\documentclass[12pt]{iopart}
\usepackage{epsfig}
\usepackage{latexsym}
\usepackage{dcolumn}
\usepackage{subfigure}
\usepackage{bm}
\usepackage{graphicx}
\usepackage{amssymb}
\usepackage{subeqn}
\usepackage[dvips]{color}

\begin{document}
\date{\today}
\title{Resistive transition in disordered superconductors with varying intergrain coupling}

\author{L. Ponta$^1$, A. Carbone$^1$ and M. Gilli$^2$ }
\address{$^1$Physics Department, Politecnico di Torino, Corso Duca degli Abruzzi 24, 10129 Torino, Italy}
\address{$^2$Electronics Department, Politecnico di Torino, Corso Duca degli Abruzzi 24, 10129 Torino, Italy}
\eads{\mailto{linda.ponta@polito.it}, \mailto{anna.carbone@polito.it}, \mailto{marco.gilli@polito.it}}

\date{\today}
\begin{abstract}
The effect of disorder is investigated in granular superconductive materials with strong and weak links. The transition is controlled by the interplay of the \emph{tunneling} $g$ and  \emph{intragrain} $g_{intr}$ conductances, which depend on the strength of the intergrain coupling. For $g \ll g_{intr}$, the transition involves first the grain boundary, while for $g \sim g_{intr}$ the transition occurs into the whole grain. The different intergrain coupling is considered by modelling the superconducting material as a disordered network of Josephson junctions. Numerical simulations show that on increasing the disorder, the resistive transition occurs for lower temperatures and the curve broadens. These features are enhanced in disordered superconductors with strong links. The different behaviour is further checked by estimating the average network resistance for weak and strong links in the framework of the effective medium approximation theory. These results may be relevant to shed light on long standing puzzles as: (i) enhancement of the superconducting transition temperature of many metals in the granular states; (ii) suppression of superconductivity in homogeneously disordered films compared to standard granular systems close to the metal-insulator transition; (iii)  enhanced  degradation of superconductivity by doping and impurities in strongly linked materials, such as magnesium diboride, compared to weakly-linked superconductors, such as cuprates.
\end{abstract}
\maketitle

\section{Introduction}
\label{Sec1}

The interplay of superconductivity and disorder has intrigued scientists for several decades \cite{Beloborodov07}.
Disorder is expected to enhance the electrical resistance, while superconductivity is associated with a zero-resistance state \cite{Efetov81}.
Bardeen, Cooper and Schrieffer explained the microscopic foundation of superconductivity in terms of pairing of electrons and the emergence of a many-body coherent macroscopic wave function \cite{Bardeer57}. Electron pairing defines a global order parameter $\Delta$ whose amplitude tends to zero by increasing temperature, current  or magnetic field thus destroying the superconducting state.
Anderson showed that weak disorder cannot lead to the destruction of the pair correlations. The transition temperature $T_c$ is insensitive to the elastic impurity scattering under the hypothesis that Coulomb interaction effects and mesoscopic fluctuations are negligible \cite{Anderson59,Balatsky06}.
However, experiments performed on thin films have demonstrated a transition from the superconducting to insulating state with increasing disorder or magnetic field.
 In sufficiently disordered metals,  these effects become important and the Anderson theorem is violated \cite{Yun06,Vinokur08,Chow98,Sambandamurthy04,Efetov80}.
\\
Studies performed on homogeneously disordered conventional materials show, upon increasing disorder,
the suppression of the superconducting critical temperature $T_c$, the enhancement of the spatial fluctuations in $\Delta$
 and the growth of the $\Delta/T_c$ ratio \cite{Sacépé08,Sondhi97,Fazio01}.
More recently, impurity effects have been investigated in unconventional $d$-wave superconductors,
with the disorder causing pair breaking and suppression of $T_c$ \cite{Dubi08,Shimizu09,Kemper09,Garg08,Spivak09,Mishra09}.
The two-gap superconductivity is also affected by disorder. Experiments in neutron-irradiated MgB$_2$ show that the two-gap
feature is evident in the temperature range above 21~K,
while the single-gap superconductivity is well established
as a bulk property, not associated with local disorder
fluctuations, when T$_c$ is lowered to 11 and 8.7~K.  The irradiation
yields samples with extremely homogeneous defect structure so that the superconducting transition
remains extremely sharp even in the heavily irradiated
samples \cite{Golubov97,Putti06}.\\
A still open issue in superconductivity is the enhancement of the critical transition temperature T$_c$ when some metals are in the granular forms rather than as a homogeneous bulk. It has been found that the enhancement is strongly dependent upon the intergrain coupling by varying pressure \cite{Osofsky01,Konig99}, with many experiments confirming this phenomenon \cite{Bond65,Tsuei74,Missell84,Dynes81,Miller88,Noak85,Kubo88}.
Suppression of superconductivity in vicinity of the metal-insulator transition has been observed in homogeneous superconductors as amorphous Au$_x$Si$_{1-x}$ and Nb$_x$Si${1-x}$ \cite{Finkelstein}. Chemical substitutions and impurities in MgB$_2$ have resulted in superconductivity degradation and broadening of the $R(T)$ curve pointing to an increasing effect of the disorder in such a strongly linked class of superconductors \cite{Li04,Larbalestier01,Xi08,Klie,Sharma,Ahn,Ahn2,Parker,Rullier,Dhalle,Wei,Matsumoto,Masui,Dou,Aksan,Cui}.
\par
Arrays of Josephson junctions with well controlled parameters are a very active field of research. As well as being of interest in their own right, they are also being used to model complex phenomena as a tool to investigate the effects of disorder in granular films \cite{Lukyanov07,Kawaguchi08,Jose94,Harris91,Haviland89,Lv09,Orr86,Ponta09,Carbone10,Brandt10,Garcia10,Wang10,Hu10}.
\par
This work is aimed at investigating the role of disorder in granular superconductors with different intergrain coupling, due to the presence of either strong or weak links.
A parameter relevant to charge-carrier transport in such materials is the dimensionless \emph{tunneling conductance} $g=G/(e^2/\hbar)$, where $G$ is the average tunneling conductance between adjacent grains and ${e^2/\hbar}$ the quantum conductance.
Films with $g\gg 1$ can be modeled as arrays of resistively shunted Josephson junctions, whose state is controlled only by the value of the normal resistance, rather than by the Josephson  and Coulomb energies which are respectively defined as $E_J=(\pi/2) g \Delta$  and $E_c=e^2/C$, with  $C$ the grain capacitance.
The tunneling of normal electrons, which additionally takes place, results in the screening of the Coulomb energy, which reduces to the effective Coulomb energy $\tilde{E_c}=\Delta/(2g)$.
By comparing the Josephson energy to the effective Coulomb energy, one can notice that $E_J$ is always larger than $\tilde{E_c}$ for $g\gg 1$.  This condition ensures the onset of the superconducting state at low temperature. Experiments show indeed that samples with the normal state conductance larger than the quantum conductance (i.e. with $g\gg 1$) always  become superconducting at low temperature.
\par
A second parameter relevant to the understanding of the behavior of different granular materials is the \emph{intragrain conductance} $g_{intr}$. For standard granular systems, the condition $g \ll g_{intr}$ holds. The intragrain region  remains in the superconducting state, with the resistive transition occurring only at the grain boundaries.
The condition $g \sim g_{intr}$ holds for tightly coupled grains, corresponding to homogenously disordered materials having comparable values of the bulk and grain boundary conductances \cite{Larbalestier01,Xi08,Li04,Lima03,Sen02,Choi04}.
\par
The different role played  by the \emph{tunneling} and \emph{intragrain conductances}  is determined  by the strength of the  coupling between the grains.
In this paper, the conditions $g \ll g_{intr}$  and $g \sim g_{intr}$ are considered in details. \par An array of Josephson junctions with different intergrain coupling and disorder degree is used to model the granular superconductor.
The different contribution of $g$ and $g_{intr}$ is accounted for by a proper circuital representation of the grain and its boundary within the network.
The study is carried out by means of a numerical simulation whose main steps are summarized in Section \ref{Model}. It is worthy of remark that the simulations reported in this work are carried out by the same numerical approach of Ref.~\cite{Ponta09}, where the different correlation shown by the current noise power spectra as a function of the intergrain coupling was investigated. The numerical results concerning the transition in weak- and strong-link networks as a function of the disorder are reported in Section \ref{NumericalResults}. The transition temperature $T_c$ is lowered and the shape of the transition curve becomes smoother by increasing the disorder. Importantly, it is found that the disorder affects more dramatically networks with strong intergrain coupling. In Section \ref{Analytical}, the results are quantitatively accounted for by estimating the resistive changes in weakly and strongly linked networks according to the effective medium approximation.

\section{Numerical model}
\label{Model}
As stated in the Introduction, the main purpose of this work is the investigation of the role of disorder in the resistive transition of granular superconductors with different intergrain coupling. The study will be carried out by adopting the numerical approach  reported in Ref.~\cite{Ponta09}, whose main steps are summarized here below.
\par
\begin{figure}
\centering
\subfigure[\label{Figure1:a}]%
{\includegraphics[width=6cm,angle=0]{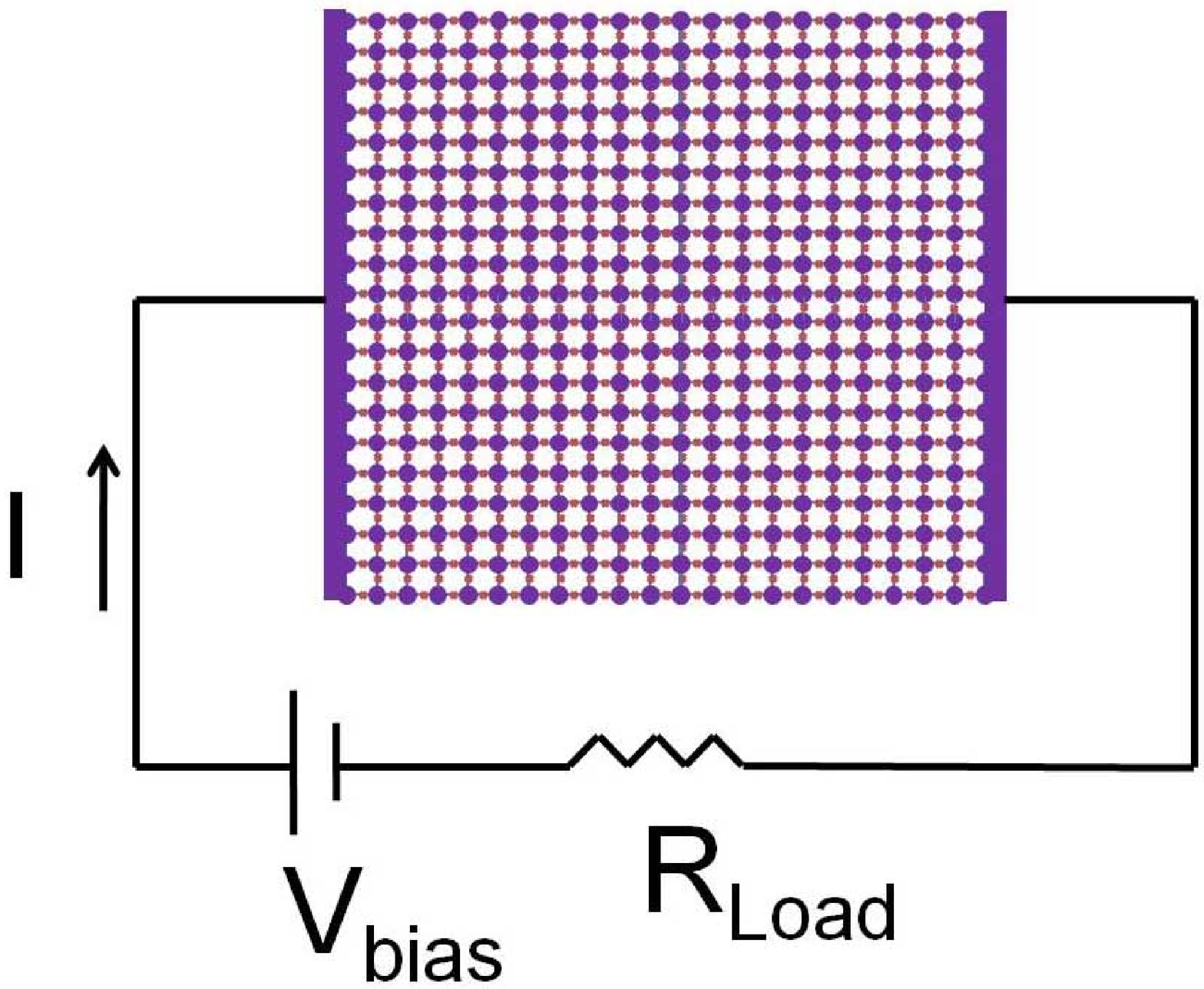}}\hspace{1em}%
\subfigure[\label{Figure1:b}]%
{\includegraphics[width=4cm,angle=0]{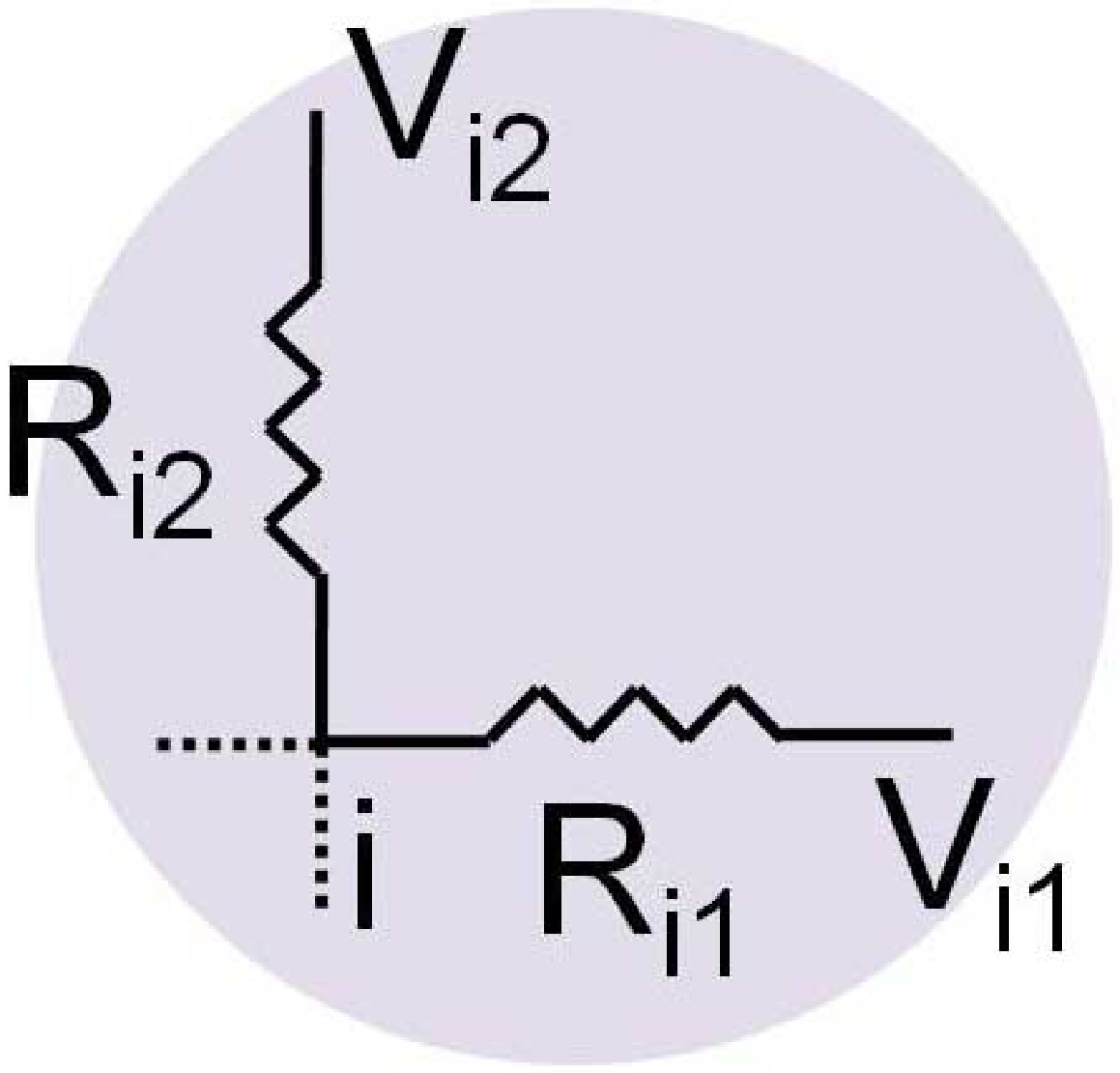}}\hspace{1em}%
\subfigure[\label{Figure1:c}]%
{\includegraphics[width=4cm,angle=0]{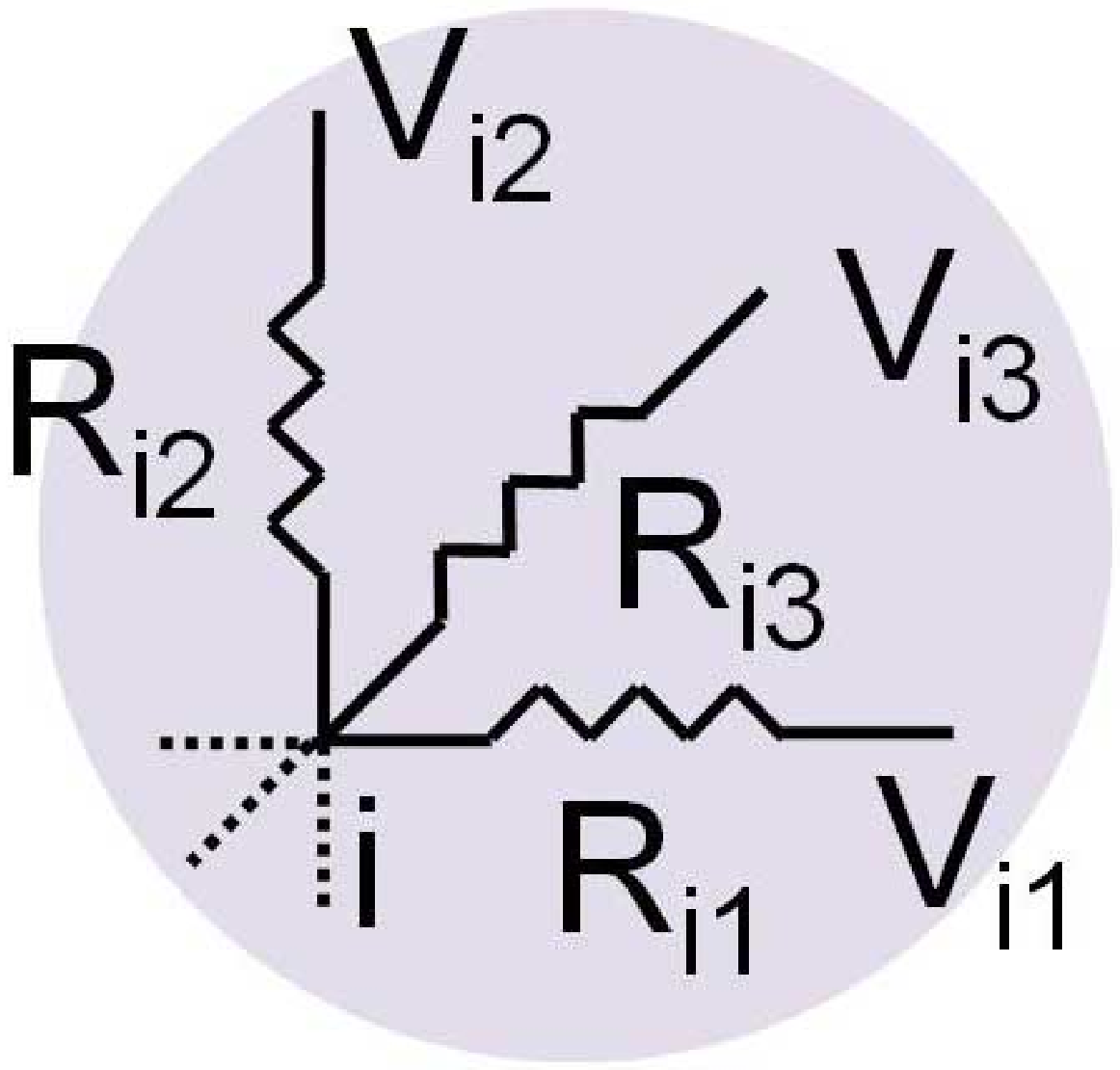}}\hspace{1em}%
\caption{ \label{Figure1} Scheme of a two-dimensional network representing the
granular superconductor (a). Grains correspond to the nodes of the arrays. Each resistor of the network behaves as a Josephson junction with characteristics schematically shown in Fig.~\ref{Figure2}. For weak-link networks (YBCO-like materials), the grains remain in the superconducting state during the first stage of the transition.  For strong-link networks (MgB$_2$-like materials), the transition is a single-stage process involving the grains. The nonlinear resistors, characterizing each grain, are respectively shown for two dimensional (b) and three dimensional (c) case.}
\end{figure}

The resistive transition is simulated by solving a system of Kirchhoff equations for a network of nonlinear resistors biased by direct current, as shown in Figure~\ref{Figure1}(a).
Two types of networks are considered for describing the different intergrain coupling.
The first type is the \emph{weak-link network} for simulating materials, whose transition occurs in two subsequent stages. First, at low temperatures, the weak-links and, then at slightly higher temperatures, the whole grain
undergoes the transition reaching the normal state. The \emph{weak-link network} is used to model the first stage of the transition occurring at the grain boundary, while the grain interior still remains superconductive.
The \emph{strong-link network} is used for modelling the transition involving the whole grain.
\par
\begin{figure}[htbp]
\centering
{\includegraphics[width=14cm,angle=0]{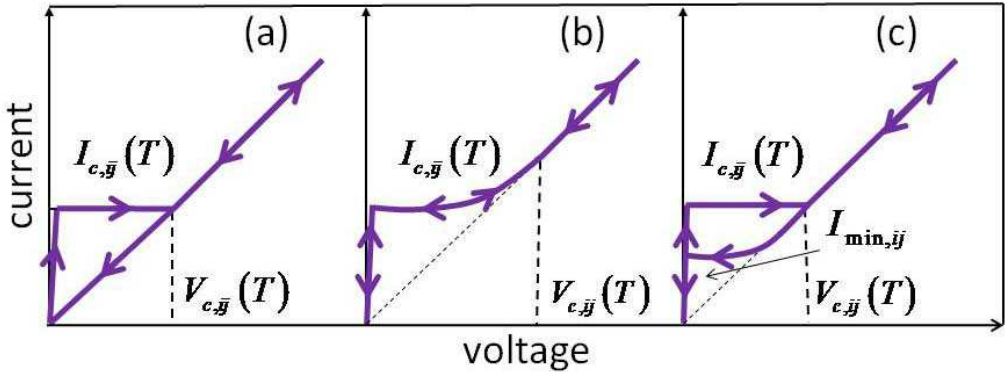}}
\caption{\label{Figure2}  Current-voltage characteristics for  underdamped (a), overdamped (b), general (c) Josephson junctions. For the general case (c), $I_{min,ij}$ depends on the Stewart-McCumber parameter $\beta_c$ and ranges from $I_{c,ij}$ and 0, where $\beta_c=\tau_{RC}/\tau_{J}$, where $\tau_{RC}$ and $\tau_{J}$ are the capacitance and Josephson time constant respectively. $\beta_c \gg 1$  (a), $\beta_c\ll1$  (b) and $\beta_c\sim 1$  (c).}
\end{figure}

Grains are represented by a couple (triple) of nonlinear resistors for two-dimensional (three-dimensional) networks of Josephson junctions as shown   respectively in Figures ~\ref{Figure1:b} and~\ref{Figure1:c}. The nonlinear resistors give a basis of independent components of the current density able to reproduce the current flowing through the grain in arbitrary directions.
 The nonlinear resistors have current-voltage characteristics as shown in Figure ~\ref{Figure2} for underdamped (a),  overdamped  (b) and  general (c) Josephson junctions. The Stewart-McCumber
 parameter $\beta_c=\tau_{RC}/\tau_{J}$, where $\tau_{RC}$ and $\tau_{J}$ are respectively the capacitance and the Josephson time constants,
  identifies the three types: $\beta_c \gg 1$ (a), $\beta_c\ll 1$ (b) and $\beta_c\sim 1$ (c).
The dependence of critical current $I_{c,{ij}}$ and magnetic field $H_{c,{ij}}$  on temperature can be written in the simplified form as:

\begin{subequations}
\begin{eqnarray}
\label{Ic}
I_{c,ij}(T)&=&I_{co,ij} \left[ 1- \left(\frac{T}{T_c}\right)^\gamma\, \right] \hspace{5pt},\\
\nonumber
\\
\label{Hc}
H_{c,ij}(T)&=&H_{co,ij} \left[ 1- \left(\frac{T}{T_c}\right)^\gamma\, \right] \hspace{5pt},
\end{eqnarray}
\end{subequations}
where $I_{co,ij}$ and $H_{co,ij}$ are respectively the low-temperature critical currents and magnetic fields
and the exponent $\gamma$ ranges approximately from $1$ to $2$ depending on material properties.
\par
The current flowing through each nonlinear resistor defines the state (superconductive, intermediate, normal) of the grain according to the current-voltage characteristics of the Josephson junction.
As already stated, the disorder is introduced in the calculations by random distribution of the critical current. The anisotropy is neglected and the same size is assumed for the grains. The reason for these simplifying assumptions is that these two features may additionally alter the network topology with a strong effect on the transition.  In particular for small grain size, the values of the critical current might be correlated in neighboring grains. Therefore, the correlation length of disorder should be taken into account by adopting a suitable spatial dependence of the critical current distribution.
 The critical current $I_{c,{ij}}(T)$  and the normal state resistance $R_{o,{ij}}$ are defined for each branch of the network. The intermediate state is characterized by the critical current $I_{c,{ij}}(T)$ and voltage drop between $0$ and $V_{\mathrm{c},{ij}}(T)$. The normal state, characterized by the resistance $R_{o,{ij}}$, is reached when the current $I$ crossing the Josephson junction exceeds $I_{c,ij}$.
The disorder is introduced by taking the critical current $I_{co,ij}$ as a random variable distributed  according to a
Gaussian distribution with mean value $I_{co}$ and standard deviation $\sigma_{I_c}=\sqrt{\sum_{ij} (I_{co,ij}-I_{co})^2/N}$.
Analogously, the disorder could be introduced by taking the critical field $H_{co,ij}$ as a random variable,
if the transition would be driven by an applied magnetic field $H$. The values of the resistances $R_{ij}$ between nodes $i$ and $j$ are taken as follows:
\begin{subequations}
\begin{eqnarray}
\label{Rijs}
R_{ij} &=& 0 \hspace{50pt}\mbox{if} ~~~~ V_{ij}\sim0 ~~~~~~~~\mbox{ (superconducting state)} \hspace{5pt}\\
\label{Riji}
R_{ij} &=& {V_{ij}}/{I_{c,ij}}\hspace{20pt}\mbox{if}~~ 0<V_{ij}<V_{c,ij}~~  \mbox{(intermediate state)} \hspace{5pt}\\
\label{Rijn}
R_{ij} &=& R_{o,ij}\hspace{35pt}\mbox{if}~~ V_{ij}>V_{c,ij} ~~~~~~~~\mbox{ (normal state)} \hspace{5pt},
\end{eqnarray}
\end{subequations}
where $V_{ij}$ is the voltage drop between nodes $i$ and $j$.
The current-voltage characteristics is used to find the value of the voltage $V_{ij}$ and current $I_{c,ij}$  by means of an iterative routine
solving the Kirchhoff equations for the network.
\par For weak-link networks, the resistance values $R_{ij}$ are calculated in a straightforward manner: the potential drops at the ends of each weak-link are compared to the potential values in the  current-voltage characteristics according to Eqs.~(\ref{Rijs},~\ref{Riji},~\ref{Rijn}). Therefore, weak-links being respectively in the superconducting, normal or intermediate state can be distinguished.
\par
For strong-link networks, the resistance values $R_{ij}$ are calculated taking into account that the voltage drop across each grain is given by:
\begin{equation}
\label{VoltageDrop}
V_i=\left[\sum_{j} V_{ij}^2\right]^{1/2} \hspace{5pt},
\end{equation}
where $V_{ij}$  corresponds to the voltage drop across each resistor $R_{ij}$ with $j=2$ or $j=3$ respectively for two- and three-dimensional arrays as shown in Figs.~\ref{Figure1:b} and  \ref{Figure1:c}.
\par
Calculations are performed iteratively. First, a tentative set of potential values is chosen for all the nodes. Then, the resistance values $R_{ij}$ are calculated by using the Josephson junction current-voltage characteristics for any resistor between  nodes $i$ and $j$. Once the $R_{ij}$ are settled, the conductance matrix with entries $G_{ij}=1/R_{ij}$  is defined and the new vector $W_1$ of the node potentials is calculated.
The set of node potentials is introduced in the iterative routine and an updated vector $W_2$ is calculated.
The iteration is repeated until the quantity $ \varepsilon_n={|W_n-W_{n-1}|}/{|W_n|}$
becomes smaller than a value $\varepsilon_{min}$ chosen to exit from the loop. The simulations are performed by varying $\varepsilon_{min}$ in the range $10^{-7} < \varepsilon_{min}< 10^{-11}$ to check that the
value of $\varepsilon_{min}$ does not appreciably change the  results.
The network resistance $R$ is then obtained by $W_n(1)/I$, where $W_n(1)$ is the potential drop at the electrodes.\\

\section{Numerical results}
\label{NumericalResults}
In this section, the  results of the numerical simulations for different degrees of disorder are reported.
It is shown that disorder affects at a different extent weak- and strong-link networks.
\par
At the beginning the network is entirely in the superconducting state (this condition is guaranteed by taking  $g\gg1$). Subsequently, the transition  is made to occur through one of these processes:
\begin{itemize}
\item the temperature  is kept constant and the bias current (or the applied magnetic field) is varied. When the  current $I_{ij}$  exceeds the critical current $I_{c,ij}$ (or the magnetic field exceeds the critical field $H_{c,ij}$), the superconductive grain evolves to the intermediate and, then, to the normal state.
\item the bias current (or the magnetic field) is kept constant and the temperature is varied. A temperature increase  causes  a decrease of critical current $I_{c,ij}$  according to Eq.~(\ref{Ic}) (or of critical field $H_{c,ij}$ according to Eq.~(\ref{Hc}))  and, ultimately, causes the transition of the grain to the intermediate and, then, to the normal state.
\end{itemize}
\par
As already stated, the disorder is modeled by assuming that the critical currents are a random variable distributed according to a Gaussian function with standard deviation  $\sigma_{I_c}$. The spread of the distribution of the critical currents determines the slope of the transition curve \cite{Markiewicz06}.
The standard deviation $\sigma_{I_c}=0$ corresponds to a fully ordered network, with all the Josephson junctions having the same critical current with
the transition occurring simultaneously all through the network. When the disorder increases ($\sigma_{I_c}$ increases),
the Josephson junctions have a wider spread of $I_{c,ij}$  and the network resistance changes more smoothly.
\par
\begin{figure}
\centering
\subfigure[\label{Figure3:a}]%
{\includegraphics[width=8cm,angle=0]{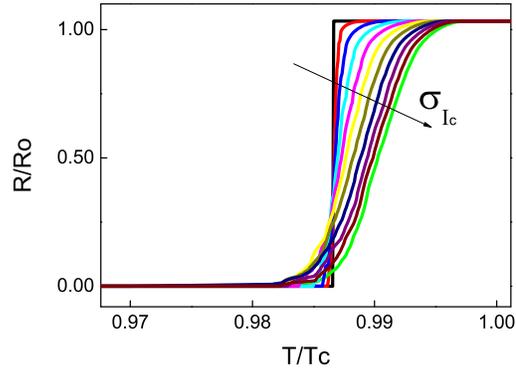}}\hspace{1em}%
\subfigure[\label{Figure3:b}]%
{\includegraphics[width=8cm,angle=0]{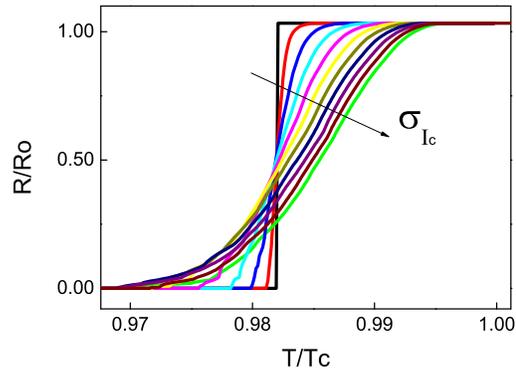}}\hspace{1em}%
\caption{\label{Figure3} Resistive transition  of two-dimensional networks with different degree of disorder at varying temperature for weak-links (a) and strong-links (b). The bias current $I$ is kept constant. The degree of disorder is varied by changing the value of the standard deviation of the critical currents $\sigma_{I_c}$  from 0 to 1 with step 0.1.}
\end{figure}

Figures~\ref{Figure3:a} and \ref{Figure3:b} show the resistive transition of the network for different values of $\sigma_{I_c}$ for weak and strong link networks respectively. The temperature increases while the external current $I$ is kept constant.
As temperature increases, the critical current $I_{c,ij}$ decreases according to Eq.~(\ref{Ic}). Links with $I_{c,ij}$ values  smaller than  $I_{ij}$ undergo the transition to the normal state.
If $\sigma_{I_c}$ is small the resistive transition is steeper. In the limit of $\sigma_{I_c}=0$ (no disorder in the network), the transition is vertical, since all the Josephson junctions become resistive for the same value of temperature. On the contrary, if $\sigma_{I_c}$ is large the resistive transition broadens since the junctions become resistive at different temperatures. This effect occurs both in weak- and strong-link network, but is enhanced in strong-link networks.
\par
\begin{figure}
\centering
\subfigure[\label{Figure4:a}]%
{\includegraphics[width=8cm,angle=0]{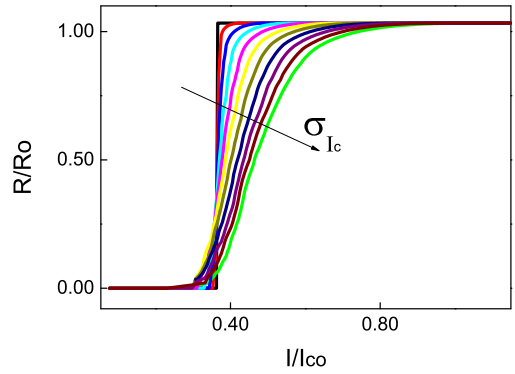}}\hspace{1em}%
\subfigure[\label{Figure4:b}]%
{\includegraphics[width=8cm,angle=0]{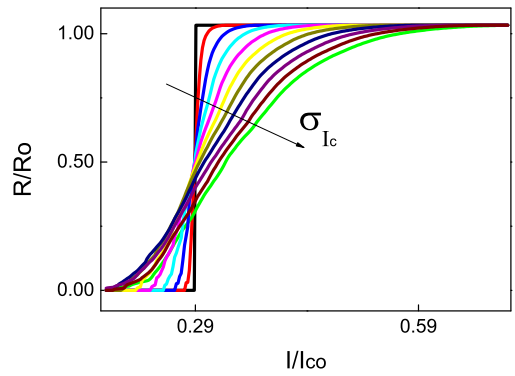}}\hspace{1em}%
\caption{\label{Figure4} Resistive transition  of a two-dimensional network with different degree of disorder at varying bias current for weak-links (a) and strong-links (b). The temperature $T$ is kept constant. The degree of disorder is varied by changing the value of the standard deviation of the critical currents $\sigma_{I_c}$ from 0 to 1 with step 0.1.}
\end{figure}

Figure ~\ref{Figure4:a} and \ref{Figure4:b} show the resistive transition when the bias current $I$ increases at constant temperature, for different values of $\sigma_{I_c}$ in  weak- and strong-link networks respectively.  When the bias current $I_{ij}$ exceeds $I_{c,ij}$, the weak links become resistive. The transition curves of Fig.~\ref{Figure4:a} and \ref{Figure4:b} exhibit a behavior similar to those of Fig.~\ref{Figure3:a} and \ref{Figure3:b}. The disorder makes the resistive transition smoother, particularly in networks with strong-links.\\

\section{Discussion}
\label{Analytical}
In this section, the results of the simulations will be discussed. One can observe that the average network resistance $R$ is determined by the elementary nonlinear resistances $R_{ij}$ between nodes $i$ and $j$.  The values of $R_{ij}$ depend on the external drive  (current, magnetic field,  temperature)  and on the intrinsic properties of the junctions.
The change of the resistance $\Delta R_{ij}$  can be expressed in terms of the external drive variation as:
\begin{equation}\label{DeltaRi1}
\Delta R_{ij}= \frac{\partial R_{ij}}{\partial I} \Delta I+ \frac{\partial R_{ij}}{\partial H} \Delta H + \frac{\partial R_{ij}}{\partial T} \Delta T \hspace{5pt}.
\end{equation}
The three terms on the right hand side of Eq.~(\ref{DeltaRi1}) can be written respectively as:
\begin{subequations}
\begin{eqnarray}
\label{dRdI}
\frac{\partial R_{ij}}{\partial I} \Delta I &=& -\frac{\partial R_{ij}}{\partial I_c} \Delta I_c \hspace{5pt},\\
\label{dRdH}
\frac{\partial R_{ij}}{\partial H} \Delta H &=& -\frac{\partial R_{ij}}{\partial H_c} \Delta H_c \hspace{5pt},\\
\label{dRdT}
\frac{\partial R_{ij}}{\partial T} \Delta T  &=& \left(\frac{\partial R_{ij}}{\partial I_c} \frac{\partial I_c}{\partial T}+ \frac{\partial R_{ij}}{\partial H_c} \frac{\partial H_c}{ \partial T}\right)\Delta T \hspace{5pt}.
\end{eqnarray}
\end{subequations}
Equations~(\ref{dRdI}) and ~(\ref{dRdH}) mean that the increase (decrease) of bias current or magnetic field acts as a decrease (increase) of critical current $I_c$ or  magnetic field $H_c$. Eq.~(\ref{dRdT}) means that the temperature affects $R_{ij}$ mostly through a decrease of the critical current and magnetic field.
By using equations~(\ref{dRdI},~\ref{dRdH},~\ref{dRdT}), with the derivatives $\partial I_c/\partial T$ and $\partial H_c/\partial T$ in Eq.~(\ref{dRdT})
 calculated by using Eqs.~(\ref{Ic},\ref{Hc}), Eq.~(\ref{DeltaRi1}) can be rewritten as:
\begin{equation}
\label{DeltaRi2}
\Delta R_{ij}= - \frac{\partial R_{ij}}{\partial I_c} \left( \Delta I_c+ \gamma \frac{I_{co}}{T_c} \Delta T \right)- \frac{\partial R_{ij}}{\partial H_c} \left( \Delta H_c+ \gamma\frac{H_{co}}{T_c} \Delta T \right) \hspace{5pt}.
\end{equation}
%
%
\par
Equation~(\ref{DeltaRi2}) relates  $\Delta R_{ij}$  to the variation of critical current $\Delta I_c$ or critical magnetic field $\Delta H_c$. One can note that
 $\Delta R_{ij}$ decreases when $\Delta I_c$  or $\Delta H_c$  increase due to the increased disorder in the array.
Hence, since the network resistance $R$ is proportional to terms varying as $\Delta R_{ij}$, the slope of the resistive transition is smoother when $\Delta I_c$  ($\Delta H_c$) increases for a given temperature increase $\Delta T$, regardless of the coupling strength between grains.
\par However, Eq.~(\ref{DeltaRi2}) cannot explain why the resistive transition becomes smoother with strong-link than with weak-links as one can notice in Fig.~\ref{Figure3} and \ref{Figure4}.
Therefore, in the following, the origin of the different behaviour exhibited by network with different intergrain coupling and same parameters of the elementary Josephson junctions, will be explained by including the effect of the different network topology.
\par
\begin{figure} 
\centering
\subfigure[\label{Figure5:a}]%
{\includegraphics[width=8cm,angle=0]{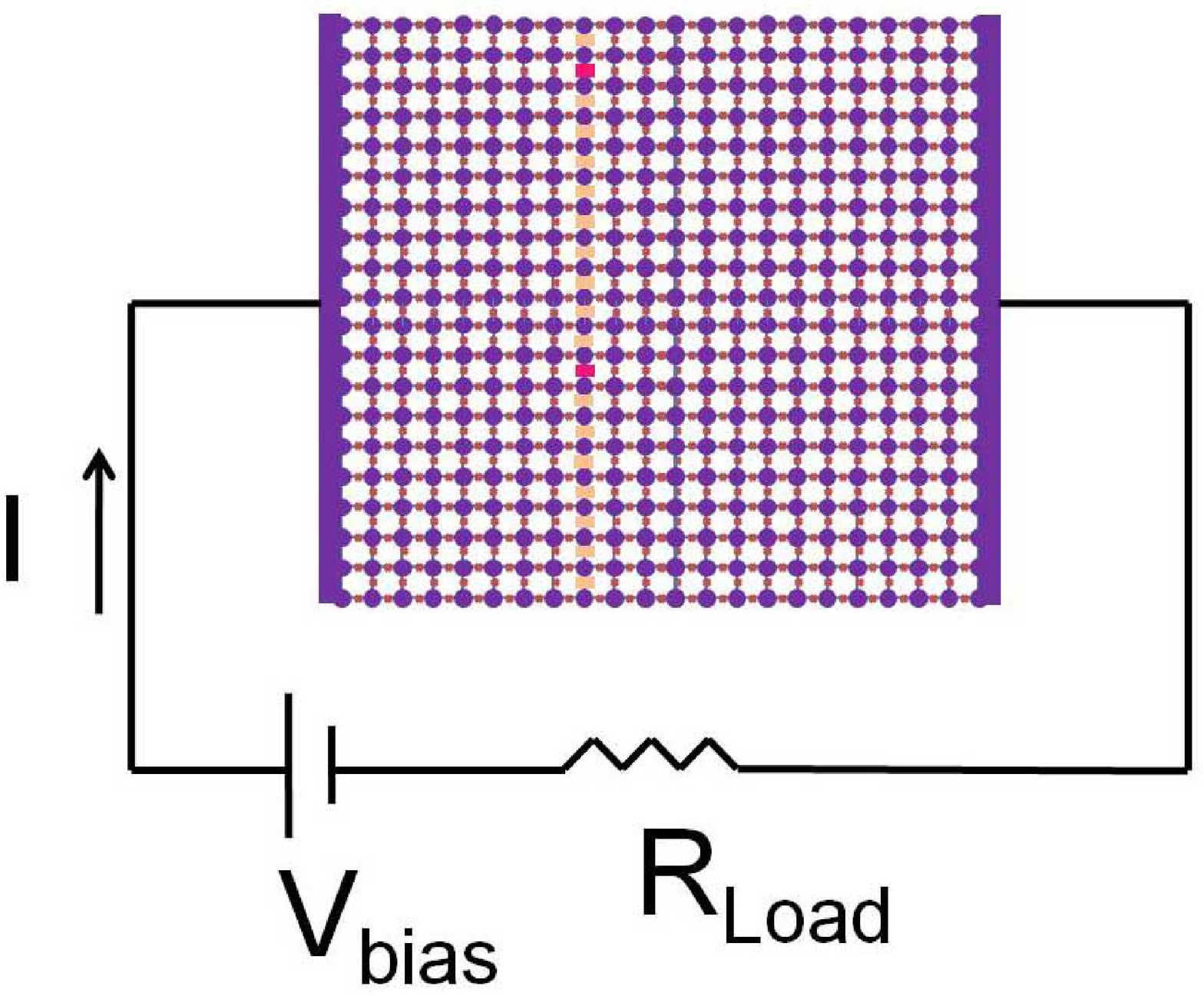}}\hspace{1em}%
\subfigure[\label{Figure5:b}]%
{\includegraphics[width=8cm,angle=0]{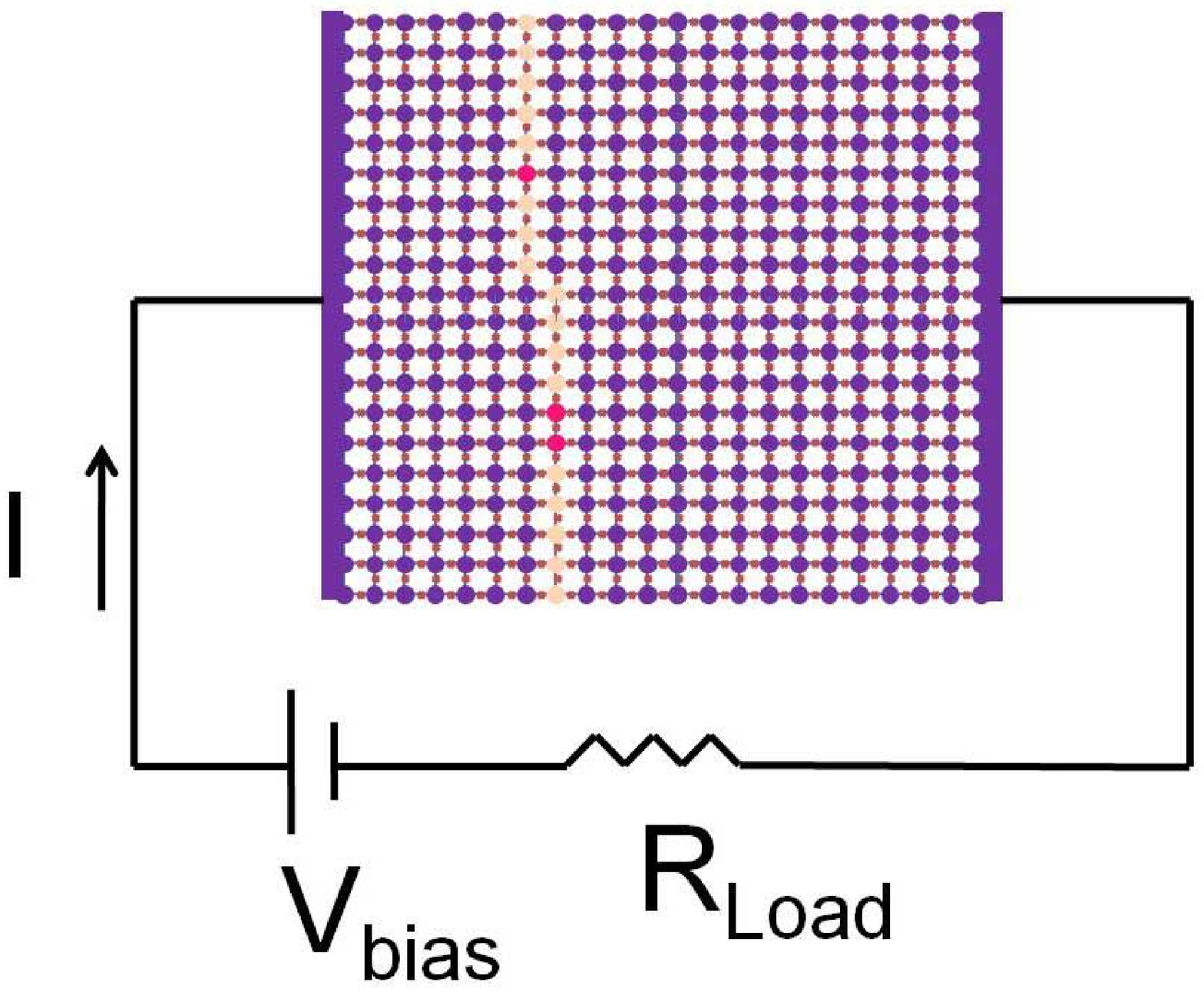}}\hspace{1em}%
\caption{\label{Figure5}  Scheme of two-dimensional networks when the first resistive layer is formed, respectively for weak-links (a) and strong-links (b).}
\end{figure}

By effect of the temperature increase, layers of weak-links or grains either in the resistive or in the intermediate state, crossing the whole film are formed  as shown is Figs.~\ref{Figure5:a} and ~\ref{Figure5:b}. The formation  of a layer corresponds to an elementary step in the network resistance. This means that, in the limit of a large number of layers, which is a reasonable condition for real granular materials, the local slope $\Delta R$ of the transition curve can be approximated by the resistance of each layer  $R_{l}$. In the remainder of this section, the resistance $R_l$ will be estimate.
Let $N_{s,l}$ label the number of weak- or strong-links  in the superconductive state before the transition of the layer. Let $N_{o,l}$ label the number of weak- or strong-links in the normal state and $N_{m,l}=N_{s,l}-N_{o,l}$ the number of weak- or strong-links in the intermediate state at a given stage of the transition of each layer.
The resistance $R_l$ can be estimated as the parallel of the normal state resistors $R_{o,{ij}}$ and the  intermediate state resistors $R_{m,{ij}}$ as:
\begin{equation}
\label{Rl}
R_{l}= \frac{R_{o,{ij}} R_{m,{ij}}}{N_{o,l}R_{m,{ij}}+N_{m,l}R_{o,{ij}}} \hspace{5pt}.
\end{equation}
The layer resistance $R_l$ depends on the ratio of the normal $N_{o,l}$ and mixed state $N_{m,l}$ resistances.
For the strong links, the voltage drop between two neighboring grains is calculated according to Eq.~(\ref{VoltageDrop})  and thus is larger than $V_{ij}$ (voltage drop across each weak-link). Therefore, since the condition given by Eq.~(\ref{Riji}) is reached earlier, the denominator of Eq.~(\ref{Rl}) is larger in layers characterized by strong-links rather than weak-links for the same degree of disorder and bias current.  This argument agrees with the fact that the resistive transition in strong-link networks occurs at temperatures lower than in weak-link networks.
\par
\begin{figure} 
\centering
\subfigure[\label{Figure6:a}]%
{\includegraphics[width=8cm,angle=0]{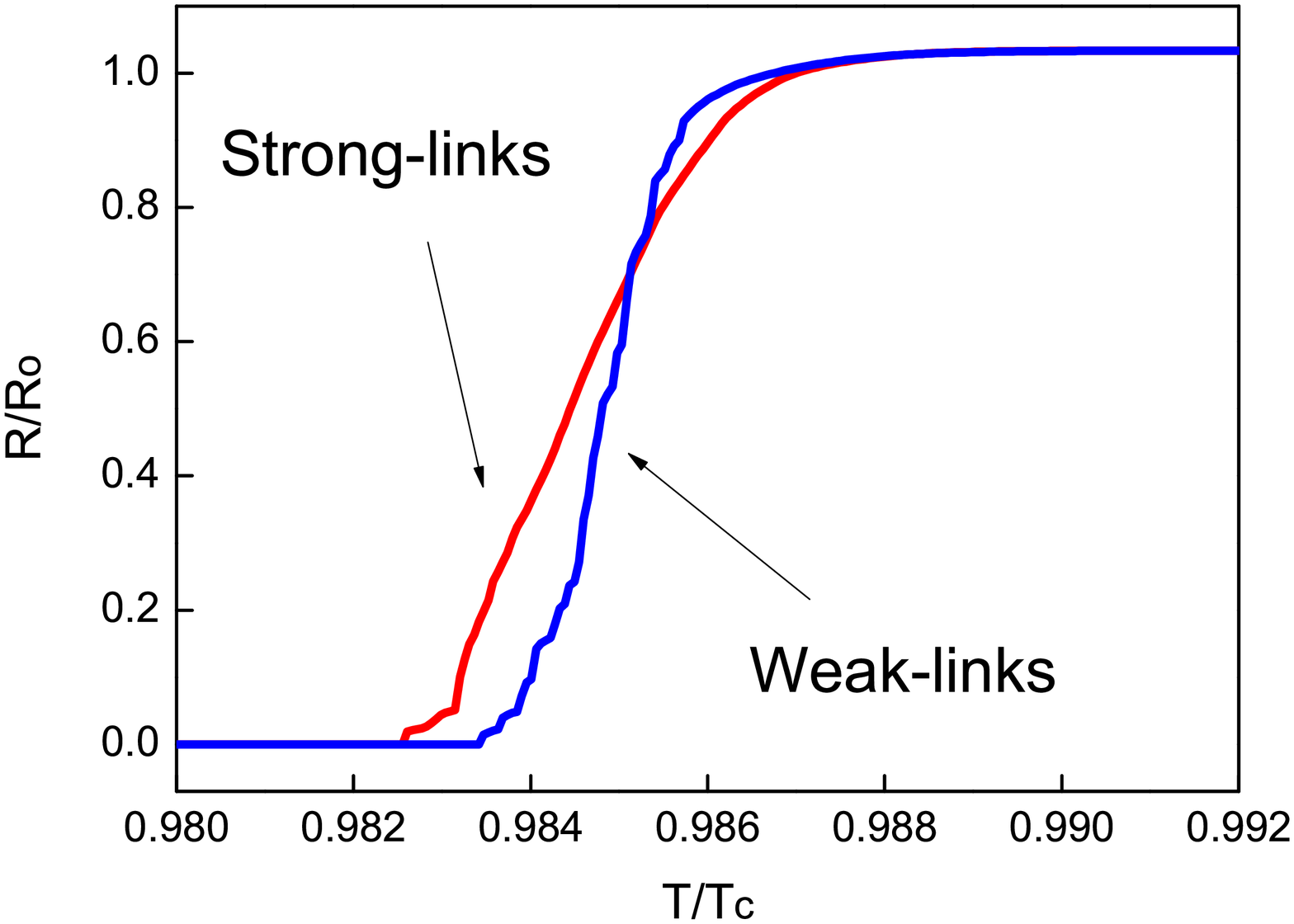}}\hspace{1em}%
\subfigure[\label{Figure6:b}]%
{\includegraphics[width=8cm,angle=0]{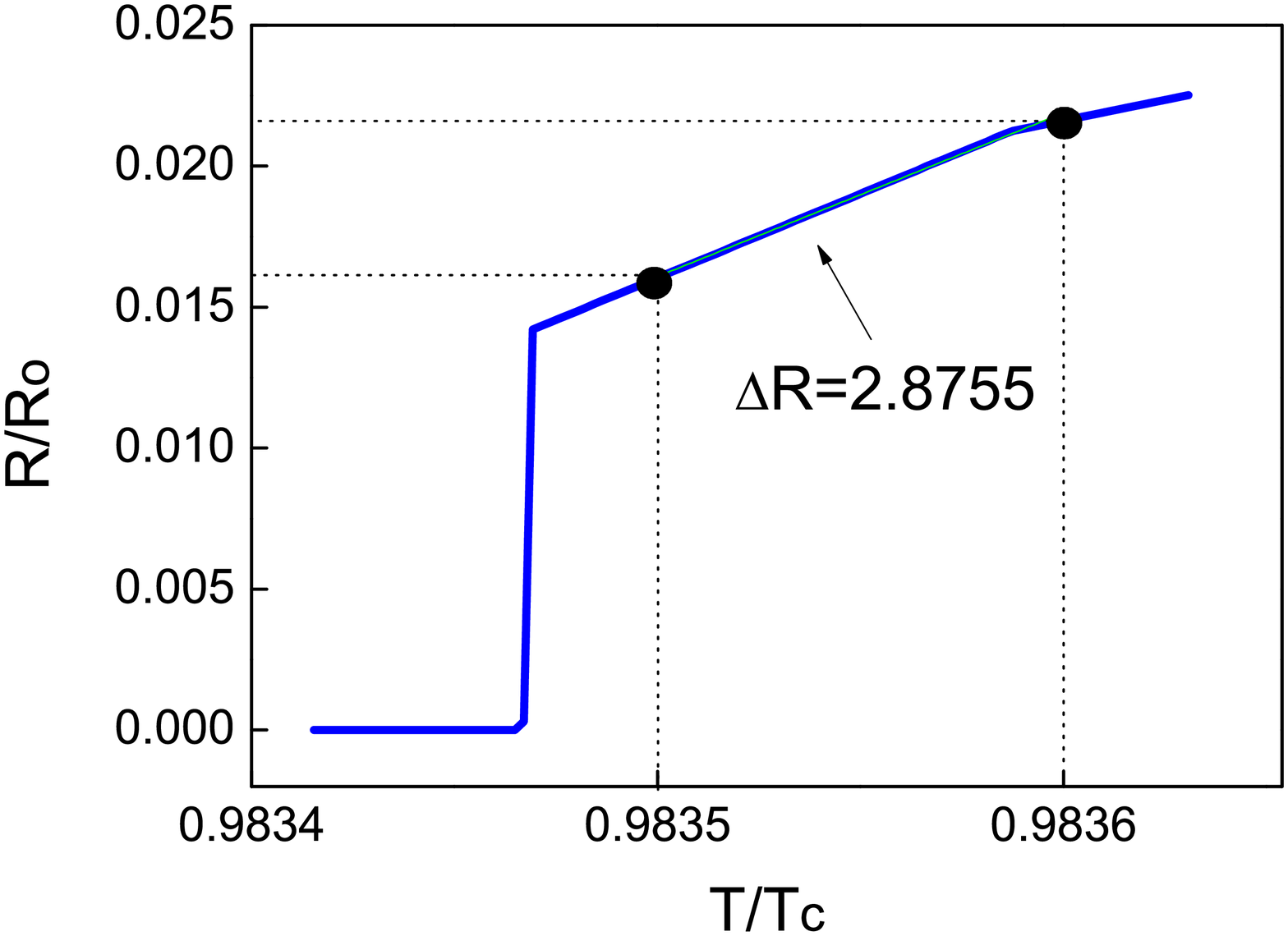}}\hspace{1em}%
\subfigure[\label{Figure6:c}]%
{\includegraphics[width=8cm,angle=0]{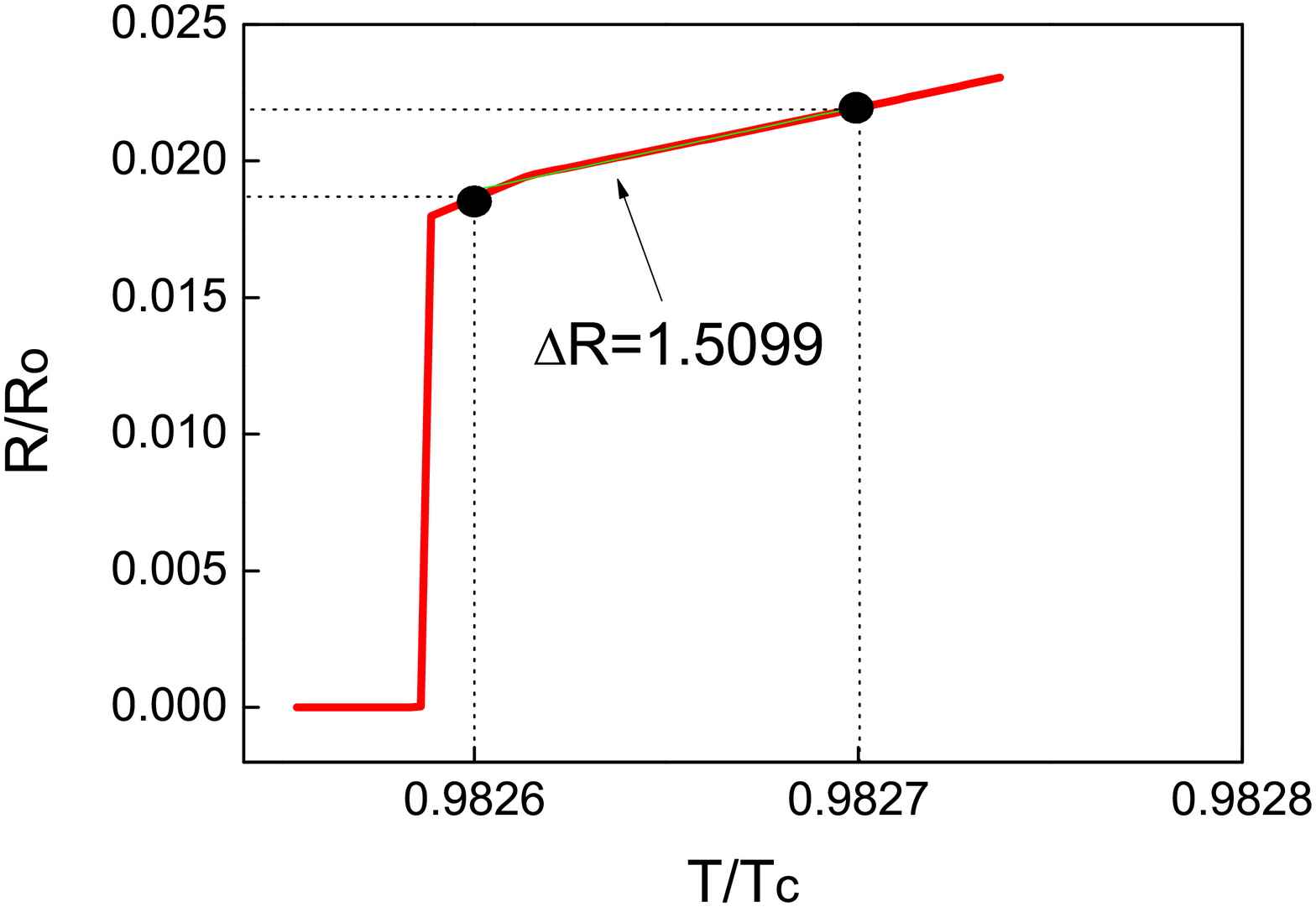}}\hspace{1em}%
\caption{\label{Figure6}  Resistive transition of two-dimensional network with weak (blue) and strong (red) links  as temperature increases. The standard deviation of the Gaussian distribution of the critical current $\sigma_{I_c}$ is equal to 0.2 for both curves. Zoom of the first step of the resistive transition in weak link (b) and in strong link  (c) network.}
\end{figure}
\begin{figure} 
\centering
\subfigure[\label{Figure7:a}]%
{\includegraphics[width=8cm,angle=0]{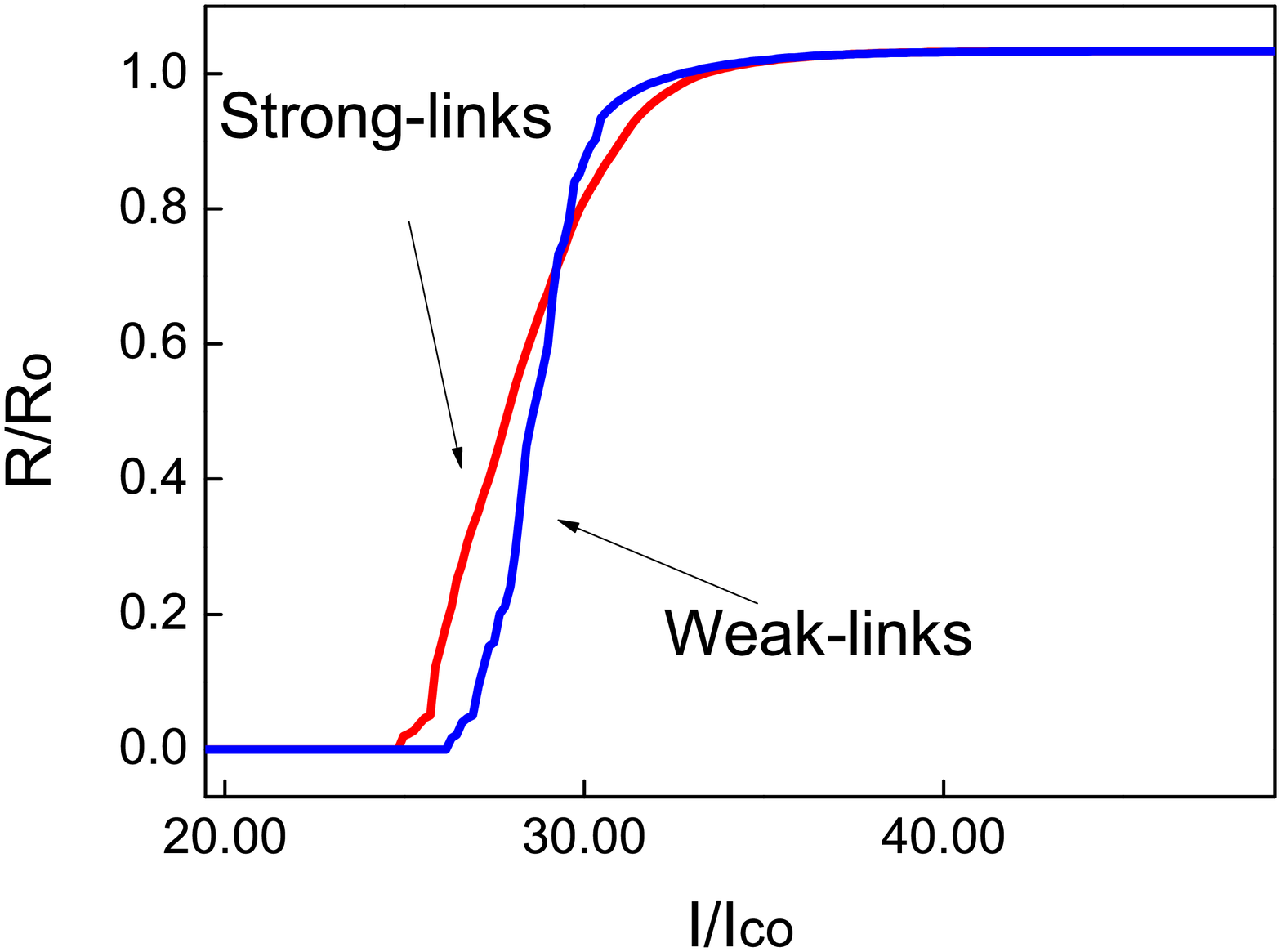}}\hspace{1em}%
\subfigure[\label{Figure7:b}]%
{\includegraphics[width=8cm,angle=0]{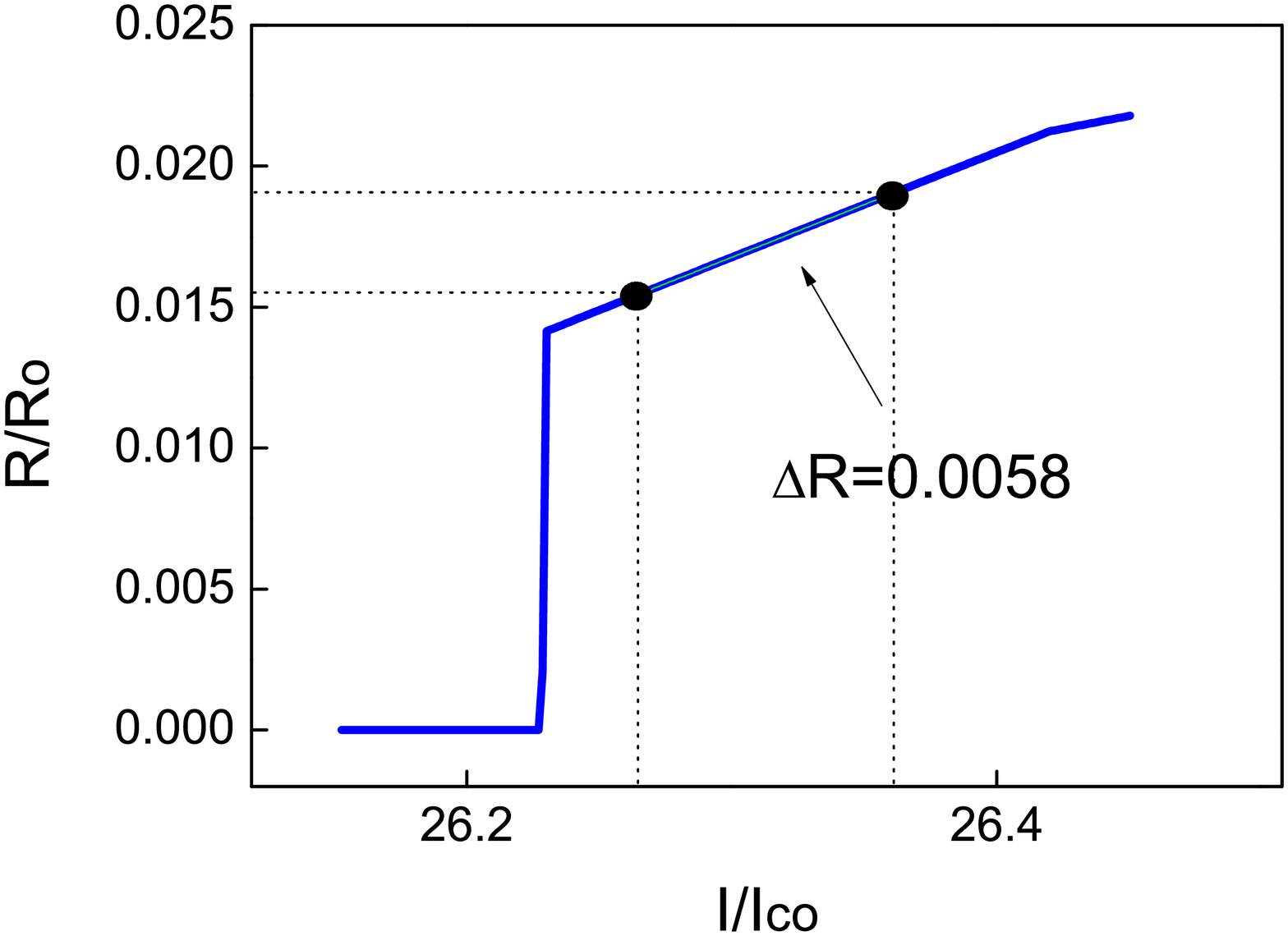}}\hspace{1em}%
\subfigure[\label{Figure7:c}]%
{\includegraphics[width=8cm,angle=0]{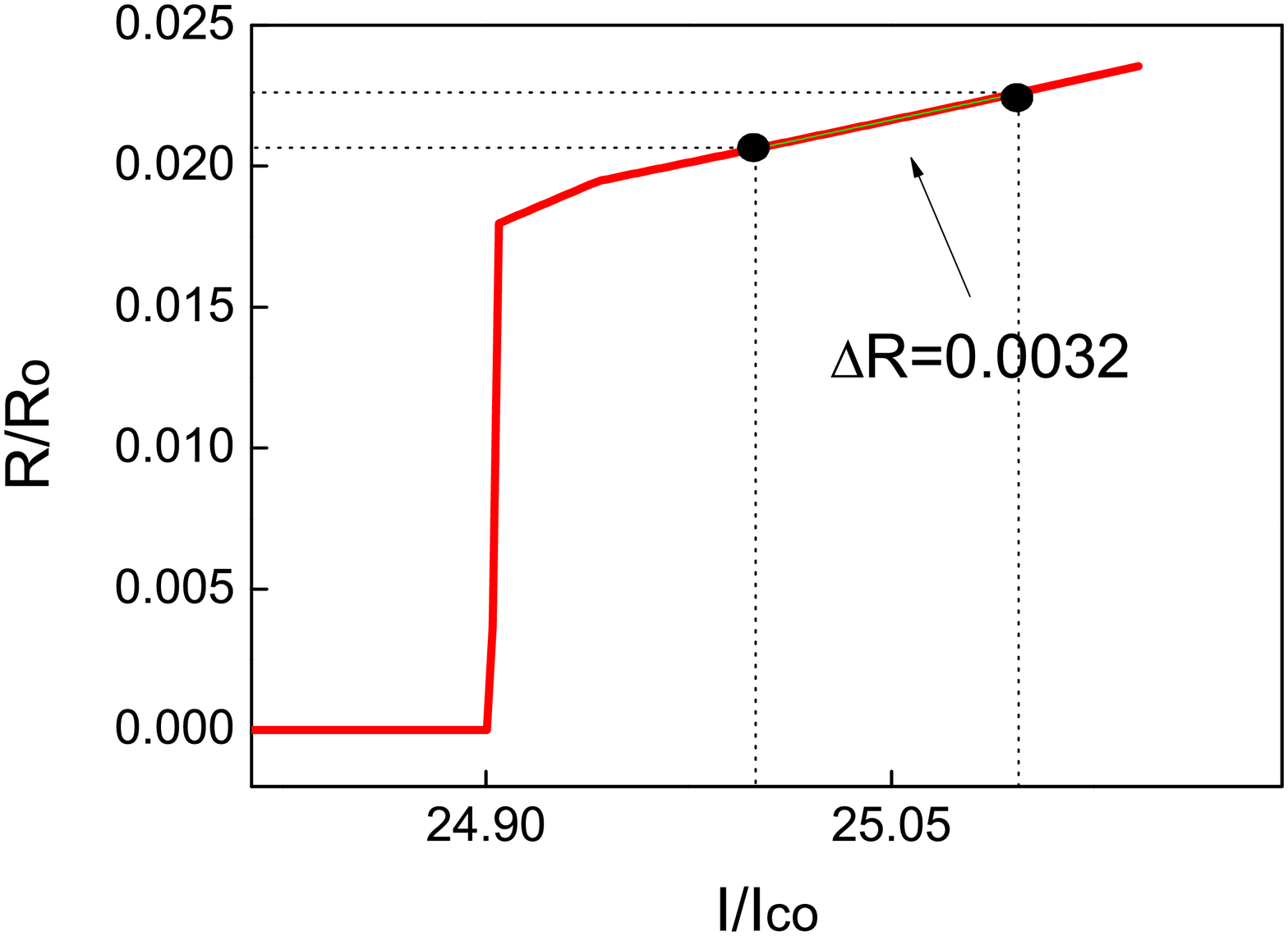}}\hspace{1em}%
\caption{\label{Figure7} Resistive transition of two-dimensional network with weak (blue) and strong  (red) links as the bias current increases. The standard deviation of the Gaussian distribution of the critical current $\sigma_{I_c}$ is equal to 0.2 for both cases. Zoom of the first step of the resistive transition in weak link  (b) and  strong link  (c)  networks.}
\end{figure}
\begin{figure}[htbp]
\centering
{\includegraphics[width=18cm,angle=0]{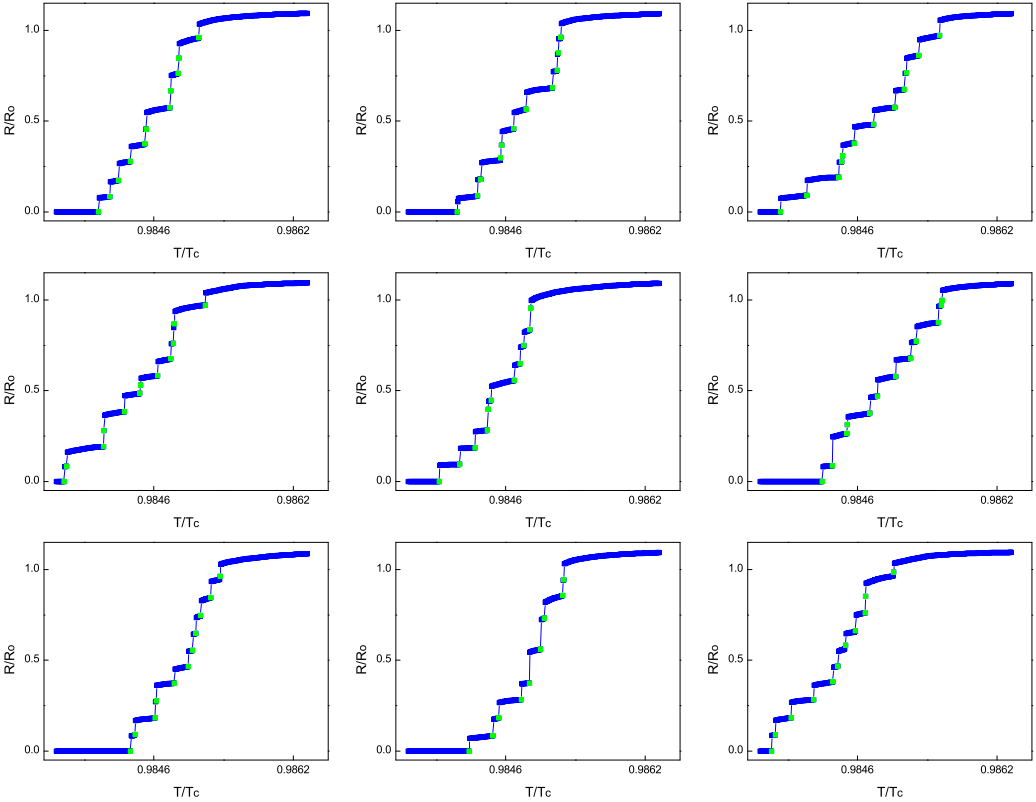}}
\caption{\label{Figure8} Resistive transitions of two-dimensional network with weak-links. Elementary resistance steps can be clearly observed. These nine curves are typical samples used for obtaining the data plotted in the histogram shown in Fig.~\ref{Figure10:a}.  }
\end{figure}
\begin{figure}[htbp]
\centering
{\includegraphics[width=18cm,angle=0]{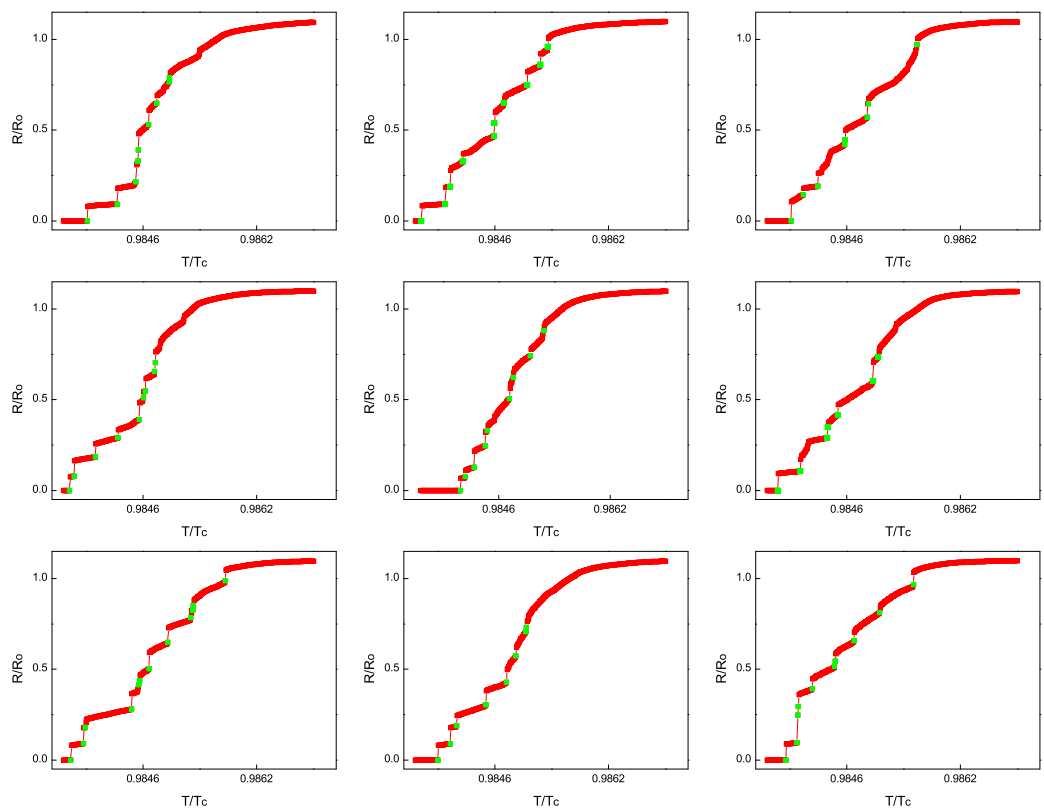}}
\caption{\label{Figure9}  Same as in Fig.~\ref{Figure8} but for strongly linked grains. The histogram  is shown in Fig.~\ref{Figure10:b}. }
\end{figure}
\begin{figure} 
\centering
\subfigure[\label{Figure10:a}]%
{\includegraphics[width=8cm,angle=0]{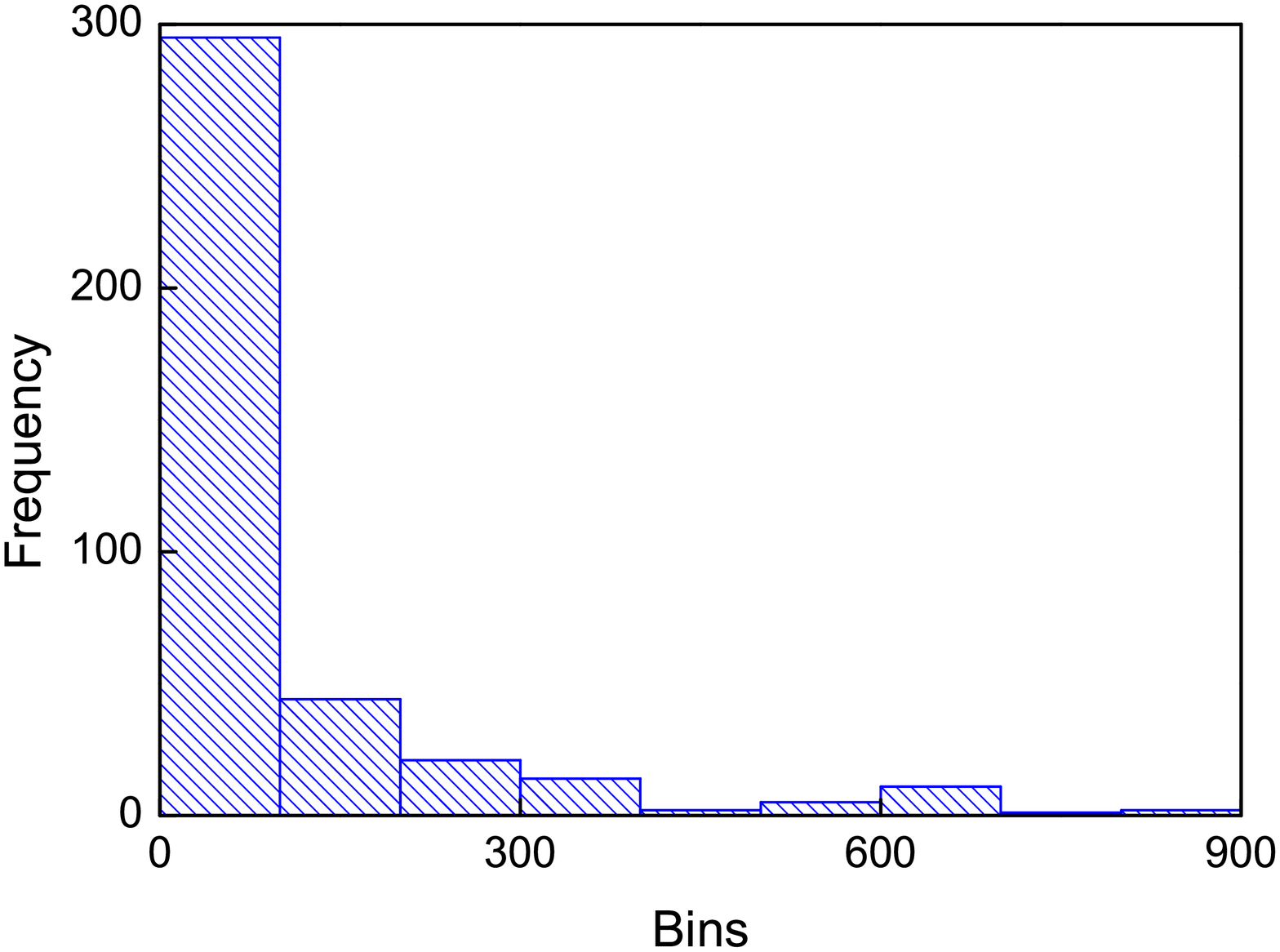}}\hspace{1em}%
\subfigure[\label{Figure10:b}]%
{\includegraphics[width=8cm,angle=0]{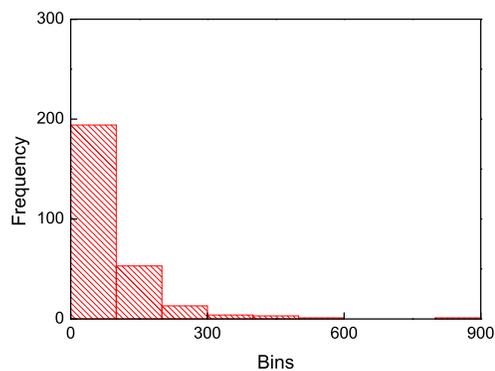}}\hspace{1em}%
\caption{\label{Figure10}  Histograms of the slopes of the elementary steps  for weak-link (a) and strong-link (b) networks. The elementary steps  have been obtained by means of transition curves similar to those shown in Figs.~\ref{Figure8} and \ref{Figure9}.}
\end{figure}

Fig.~\ref{Figure6:a} shows the transition curves  in weak- and strong-link networks with the same parameters. The slope is smaller for strong-link than for weak-link networks, consistently with the fact that the denominator of Eq.~(\ref{Rl}) is larger  and thus $\Delta R \approx R_{l}$ is smaller.
Furthermore, one can notice by comparing Figs.~\ref{Figure6:b}, \ref{Figure6:c}  that the steps are higher for strong-links.
This behavior has been confirmed by several runs of the transition simulations.
 Figs.~\ref{Figure8} and ~\ref{Figure9} show nine samples of the resistive transition for weak and strong links respectively. One can clearly notice the different shape of the elementary steps. By implementing an automatic detection process of the steps endpoints, the elementary derivatives can be estimated. Fig.~\ref{Figure10} show the histograms  of about 400 step slopes for weak-link (a) and strong-link (b) networks. This statistical analysis can be used for estimating an average value of the step slopes. The average ratio between derivatives for weak and strong links ranges between 1.3 and 2.
A similar behavior is exhibited by the transition caused by current increase as shown in Fig.~\ref{Figure7}.   To explain this issue, the elementary
resistance between two neighboring sites $i$ and $j$ will be now estimated  by using \cite{Miller60,Ambegaokar71,Shklovskii84,Strelniker07a}:
\begin{equation}
\label{Rijg1}
R_{ij} = R_o \exp\left( \frac{\epsilon_{ij}}{k_BT} + \frac{r_{ij}}{r_o}\right),
\end{equation}
where $R_o=Tk_B/ (e^2 \gamma_{ij}^0)$,$\gamma_{ij}^0$ is a rate constant related to the
electron-phonon interaction ($k_B/e^2 \gamma_{ij}^0\sim 1$),
$r_{ij}$ is the distance between two sites,
$r_o$ is the scale over which the wave function decays outside the grain,
$\epsilon_{ij}$  is the zero field activation energy given by
$\epsilon_{ij}= \Delta_{ij}(T)+E_{c,ij}$, with $E_{c,ij}=\beta e^2 r_{ij}/ (\pi \epsilon_o \epsilon d^2)$ the Coulomb energy and $d$ the mean size of grain.
Therefore,  Eq.~(\ref{Rijg1}) can be written as:
\begin{equation}\label{Rijg2}
R_{ij} = R_o \exp\left(\frac{\Delta_{ij}(T)}{k_B T}+ \frac{r_{ij}}{r_o^*} \right) \hspace{5pt},
\end{equation}
with $1/r_o^* = \left[1/r_o+\beta e^2 /(2 \pi \epsilon_o \epsilon d^2 k_B T)\right]$.
In Eq.~(\ref{Rijg2}), the resistance $R_{ij}$ explicitly depends on the quantity  $r_{ij}$, which is the effective distance seen by an electron flowing from grain $i$ to $j$.
The effective distance $r_{ij}$ is different for electrons flowing either in weak- or strong-link networks. Such a difference can be estimated
by taking into account that at constant current the voltage drop $V_{ij}$ is proportional to $r_{ij}$.
The voltage drop for the strong-link case is given by Eq.~(\ref{VoltageDrop}).  A reduction  of a factor $V_{ij}/[\sum V_{ij}^2]^{1/2}$ of the distance $r_{ij}$
in comparison to the weak-link case should be correspondingly taken into account. In the simplest case of isotropic spherical grains, $V_{ij}$ is the same in any direction, thus the reduction factor
is $1/\sqrt{2}$ or $1/\sqrt{3}$ respectively for two- and three-dimensional networks.
\par
\begin{figure} 
\centering
\subfigure[\label{Figure11:a}]%
{\includegraphics[width=7cm,angle=0]{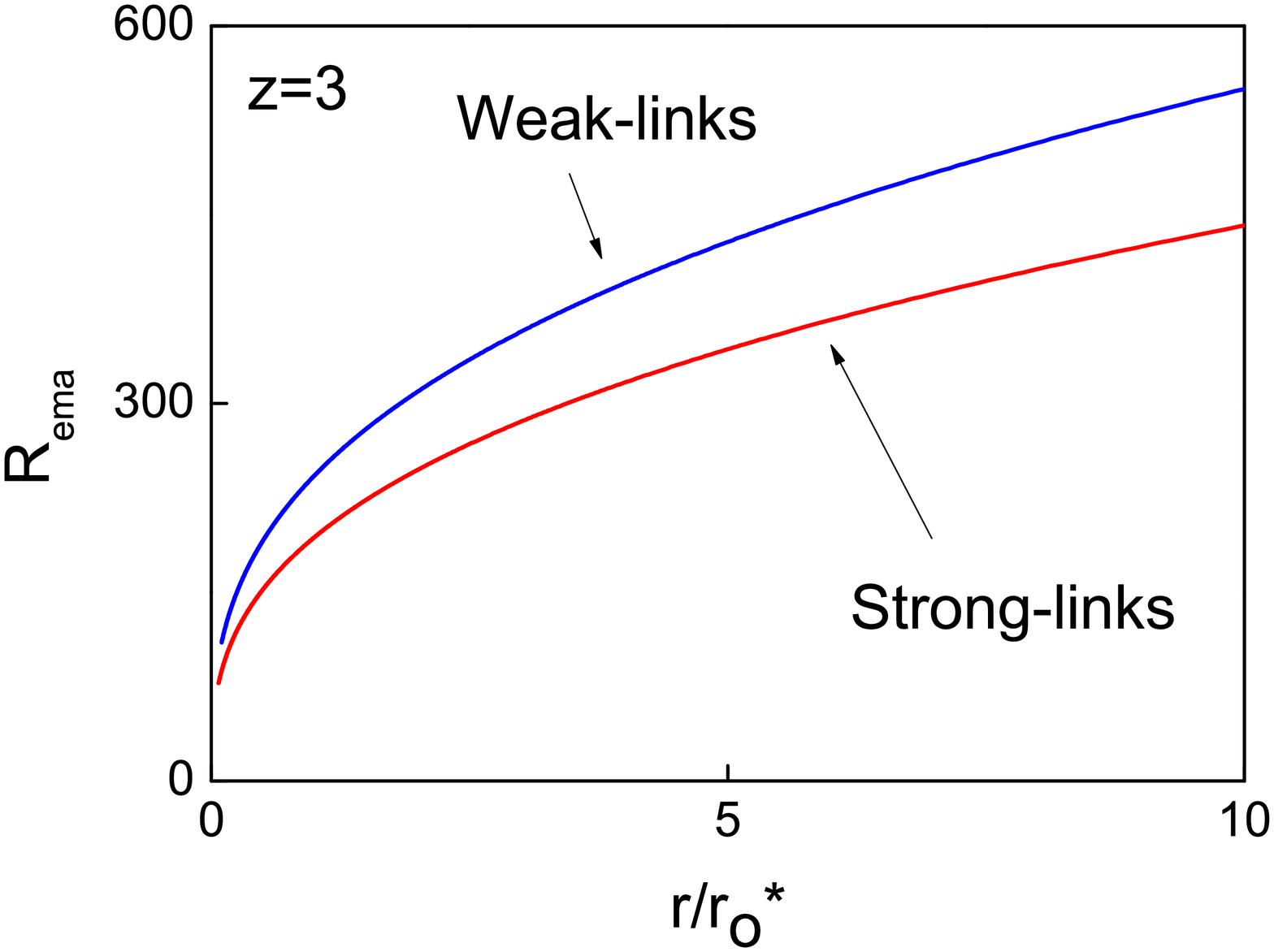}}\hspace{1em}%
\subfigure[\label{Figure11:b}]%
{\includegraphics[width=7cm,angle=0]{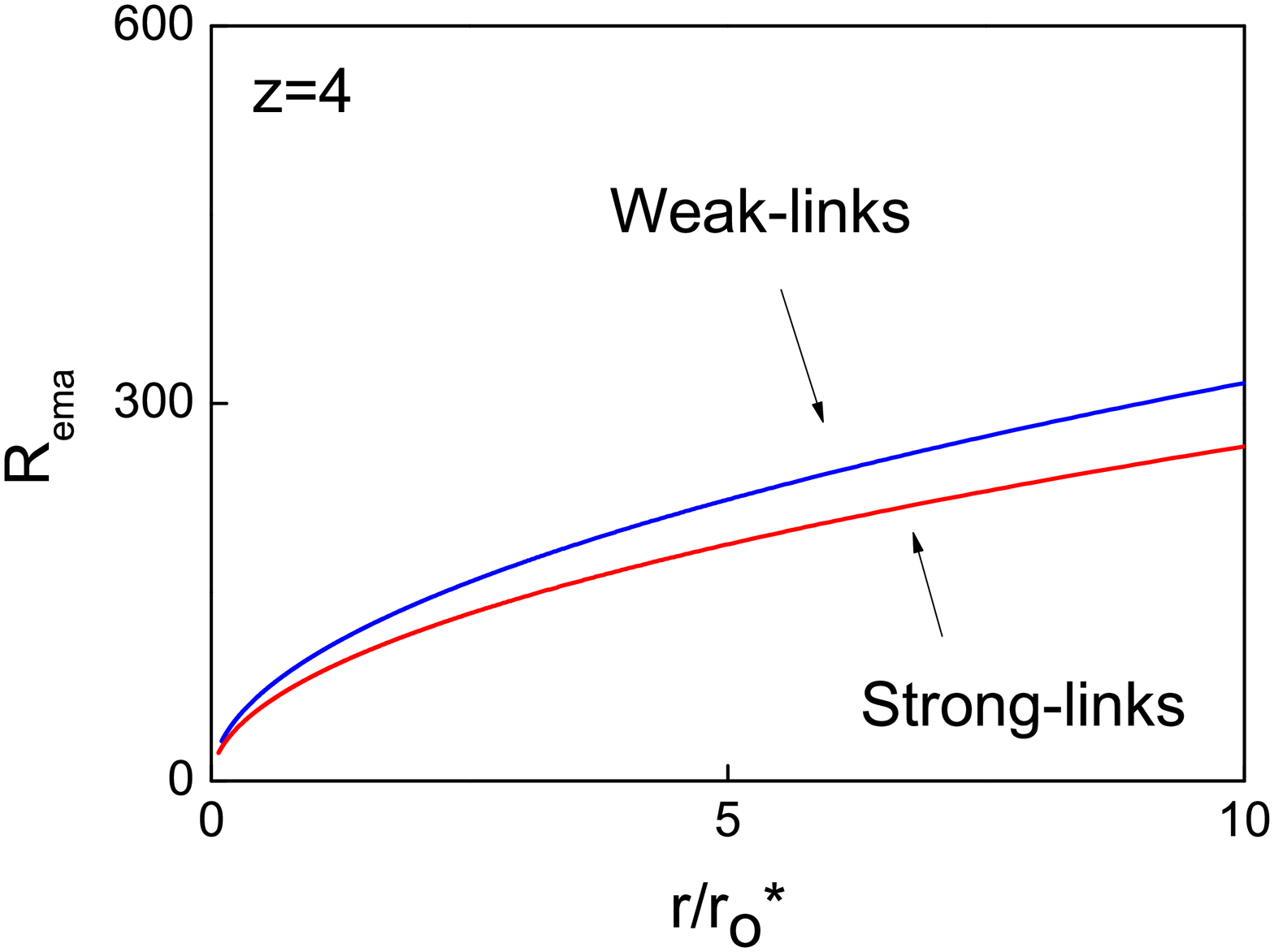}}\hspace{1em}%
\subfigure[\label{Figure11:c}]%
{\includegraphics[width=7cm,angle=0]{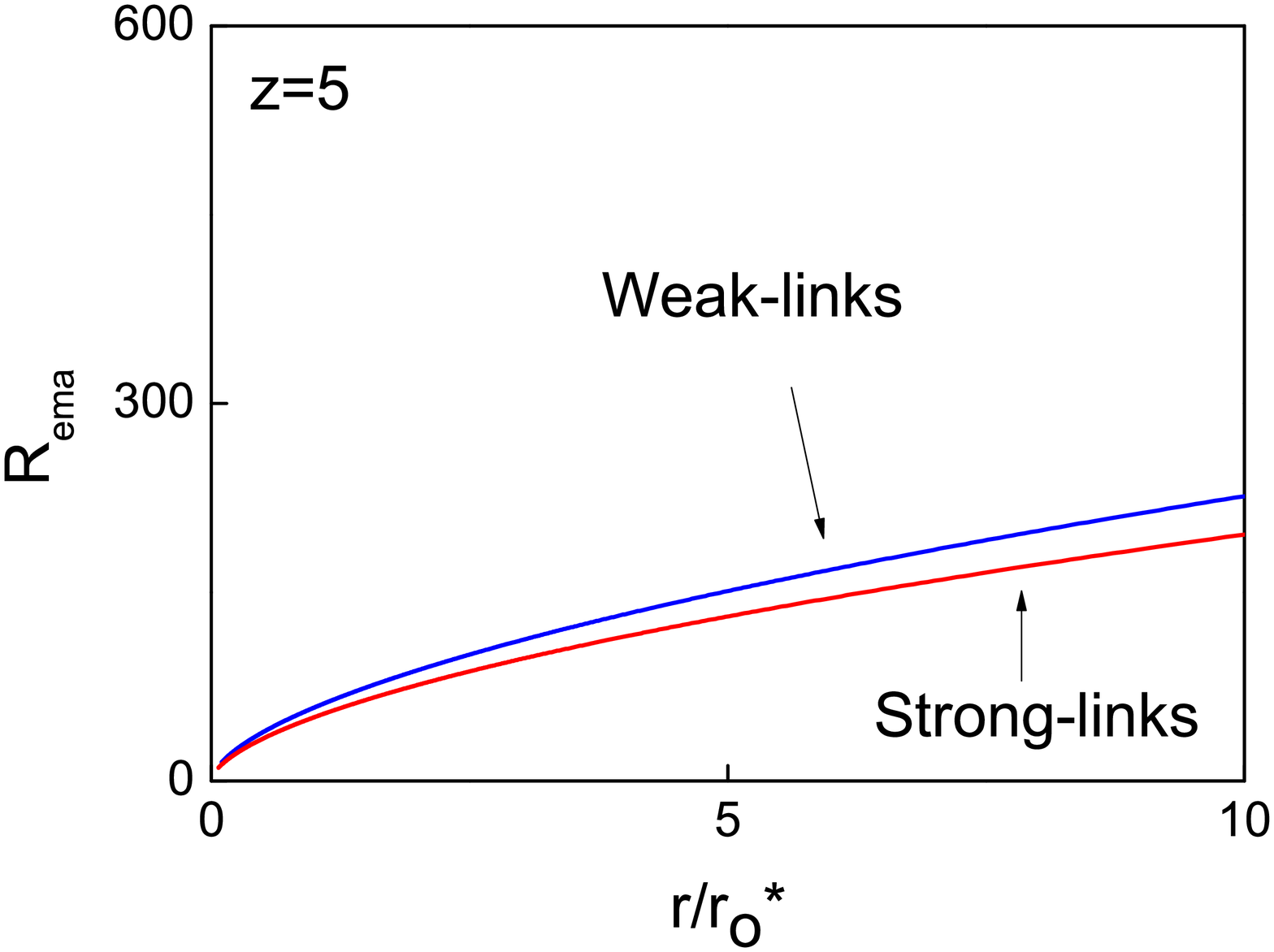}}\hspace{1em}%
\subfigure[\label{Figure11:d}]%
{\includegraphics[width=7cm,angle=0]{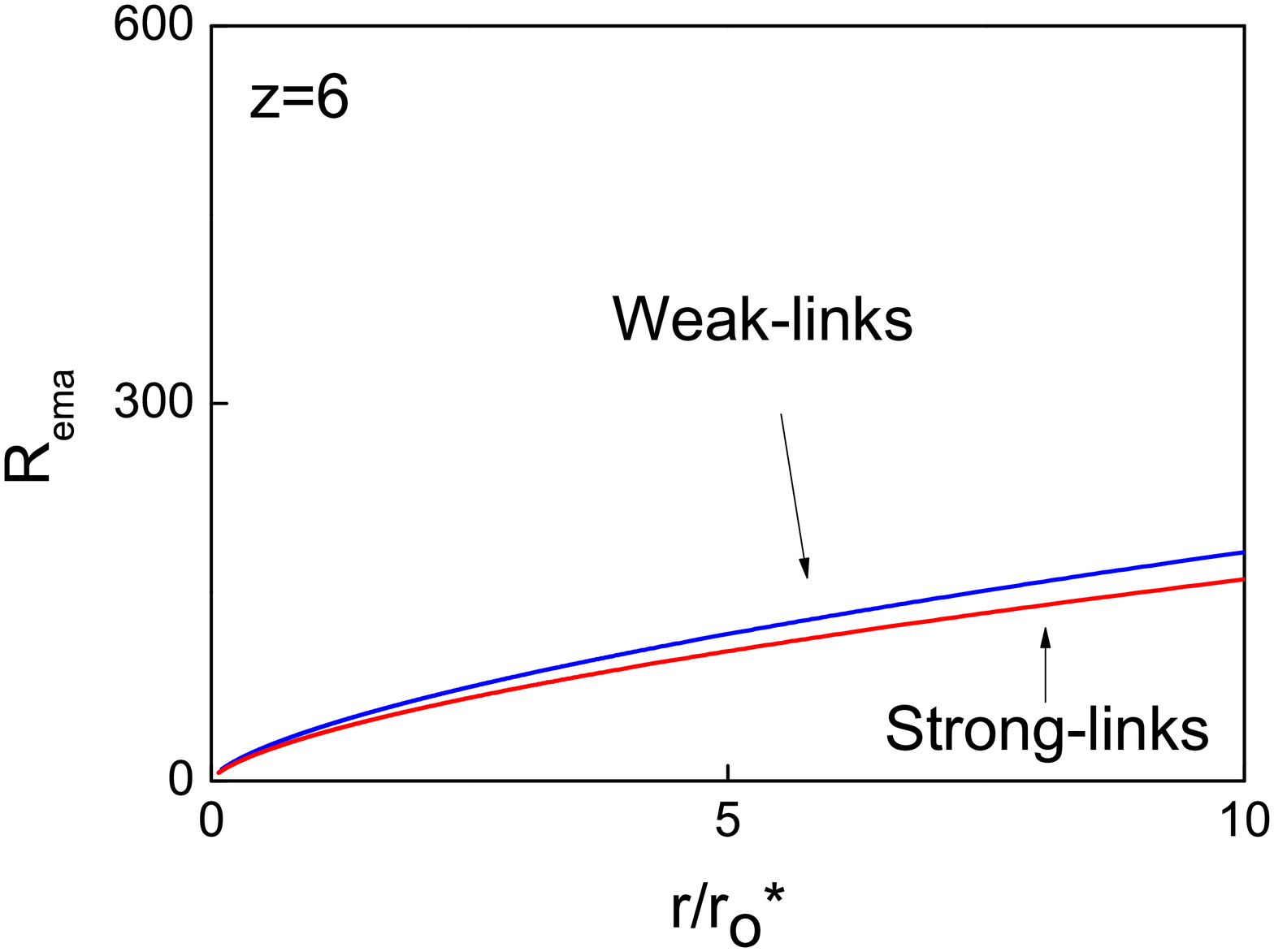}}\hspace{1em}%
\caption{\label{Figure11} Average network resistance $R_{ema}$ calculated according to the effective medium approach for $\mathrm{z}=3$ (a), $\mathrm{z}=4$ (b), $\mathrm{z}=5$ (c), $\mathrm{z}=6$ (d). Red curves refer to strongly coupled grains. Blue curves refer to weakly coupled grains. One can observe that the average resistance is smaller for strongly coupled networks for all the $\mathrm{z}$ values.}
\end{figure}

By using  the effective medium approximation \cite{Kirkpatrick73}, the average conductance $G_{ema}$ of the network can be calculated as follows:
\begin{equation}\label{kirkpatrick}
\int dG_{ij} f(G_{ij}) \frac{G_{ema}-G_{ij}}{G_{ij}+(z/2-1) G_{ema} } = 0 \hspace{5pt}.
\end{equation}
where $z$ is the number of bonds at each node of the network, $G_{ij}=1/R_{ij}$ and  $f(G_{ij})$ is the probability distribution function of the elementary conductance values $G_{ij}$. If the values $G_{ij}$ are continuously distributed according to the uniform function $f(G_{ij}) \propto 1/ G_{ij}$, the average conductance is given by:
\begin{equation}\label{Gema}
G_{ema} = G_2 \frac{ \left[\left(\frac{G_1}{G_2}\right)^{2/z}-\frac{G_1}{G_2}\right]}{\left(\frac{z}{2}-1\right)\left[1-\left(\frac{G_1}{G_2}\right)^{2/z}\right]} \hspace{5pt}.
\end{equation}
The average conductance $G_{ema}$ varies as $G_2$ times a factor depending to the ratio $G_1/G_2$. The ratio $G_1/G_2$ is independent of the intergrain coupling contrarily to $G_2$.
Therefore one can observe that the average conductance $G_{ema}$ increases as $G_2$ increases with the coupling strength.
According to presented model of the intergrain coupling, the value of the conductance $G_2$ in case of strong and weak-link networks differs of the factor $V_{ij}/[\sum V_{ij}^2]^{1/2}$. The average resistance $R_{ema}=1/G_{ema}$ is plotted in Fig.~\ref{Figure11}. It is worth noting that the resistance $R_{ema}$ for the strong-link case is always smaller then for the weak-link case as expected from the simulations.
The presented discussion could be useful to explain existing experimental observations in
granular materials that is very hard to understand with conventional mechanisms . \cite{Bond65,Tsuei74,Missell84,Dynes81,Miller88,Noak85,Kubo88}.
\section{Conclusions}
 The  effect of disorder  has been studied in superconductors with different strength of the intergrain coupling. The superconductor has been modeled as an array of Josephson junctions, numerically solved by using Kirchhoff equations.
The analysis shows that, on varying the external drive (temperature, current, magnetic field), the resistive transition occurs for lower $T_c$  and the $R(T)$ curve broadens by increasing the disorder through a stepwise process. Importantly, it is found that the effect of disorder is more dramatic when the network simulates strongly rather than weakly coupled granular superconductors. The approach used and the results obtained in this work might add useful clues  on the issue of the wide variability of critical temperature transition observed in real granular materials. It has been indeed observed an increase of critical temperature in compacted metallic powder compared to bulk samples of the same material. A strong anticorrelation between the critical temperature $T_c$ enhancement and the value of metallic conductivity has been observed indicating that a major role is played by the electron-electron interaction which acts by suppression of the superconductivity \cite{Osofsky01,Konig99}.\\
Chemical substitutions for Mg or B have been attempted to vary the superconducting transition temperature of MgB$_2$. Most of the substitutions have produced a depression of T$_c$ and broadening of the $R(T)$ curve contrarily to what observed in cuprates in which replacement  of La by Y raises T$_c$ from 35K to 93K and sharpens the transition curve. It has been suggested that the two-band nature of MgB$_2$ can result in an unusual behavior of its resistivity and T$_c$ as the material changes from the clean to dirty limits \cite{Li04,Larbalestier01,Xi08,Ahn,Ahn2}. The suppression/enhancement of T$_c$ is related to the competing effect of electron-electron and electron-phonon interaction which in their turn depend on size and radii of the compound and constituents. Intergrain and intragrain effects of disorder have been observed. Formation of magnesium or boron oxides result in poorly connected grains with an increase of intergrain resistivity and decrease of critical current density \cite{Klie,Sharma}. At the same time, these oxides might migrate within the grains themselves increasing intragrain resistivity and flux pinning. Other impurities such as silicon, carbon, copper greatly affect critical current, temperature and resistivity \cite{Parker,Rullier,Dhalle,Wei,Matsumoto,Masui,Dou,Aksan}. Critical temperature degradation and broadening of the $R(T)/R_o$ curve has been also observed in MgB$_2$ film by exposure to water\cite{Cui}.
The general feature of these experiments is that degradation of superconductivity seem to be related to the enhanced role of electron-electron interaction and impurity scattering  in homogeneous metallic-like superconductors compared to the standard granular ones, i.e. that class of material whose intergranular conductance $g$ is much smaller than the intragranular conductance $g_{intr}$. The dominant effect of the electron-electron interaction is taken into account in the present model by introducing a suitable circuital coupling among grains.

\section{Acknowledgements}
The Istituto Superiore Mario Boella is gratefully acknowledged for financial support.

\section{References}
\bibliographystyle{unsrt}
\bibliography{SuperconduttoriReferences}

\newpage

\end{document}